\documentclass[twocolumn, tighten]{aastex61}
\usepackage{graphicx}
\usepackage{amsmath, amssymb, amsfonts}
\usepackage{epstopdf}
\usepackage{natbib}
\usepackage{subfigure}
\usepackage{xspace}

\renewcommand{\micron}{$\mu$m\xspace}
\newcommand{\spitzer}{\textit{Spitzer}\xspace}
\newcommand{\spuds}{SpUDS\xspace}

\makeatletter
\newcommand*\mysize{%
  \@setfontsize\mysize{6}{6.0}%
}
\makeatother

\shortauthors{Paterno-Mahler et al.}
\shorttitle{The COBRA Survey}
\begin{document}

\title{The High-Redshift Clusters Occupied by Bent Radio AGN (COBRA) Survey: {The \spitzer Catalog}}

\correspondingauthor{R. Paterno-Mahler}
\email{rachelpm@umich.edu}

\author{R. Paterno-Mahler}
\affil{Astronomy Department and Institute for Astrophysical Research, Boston University, 725 Commonwealth Ave, Boston, MA 02215, USA}
\affil{Department of Astronomy, University of Michigan, 1085 South University Drive, Ann Arbor, MI 48109, USA}

\author{E.~L. Blanton}
\affil{Astronomy Department and Institute for Astrophysical Research, Boston University, 725 Commonwealth Ave, Boston, MA 02215, USA}

\author{M. Brodwin}
\affil{Department of Physics and Astronomy, University of Missouri-Kansas City, 5110 Rockhill Road, Kansas City, MO 64110, USA}

\author{M.~L.~N. Ashby}
\affil{Harvard-Smithsonian Center for Astrophysics, 60 Garden Street, Cambridge, MA 02138, USA}

\author{E. Golden-Marx}
\affil{Astronomy Department and Institute for Astrophysical Research, Boston University, 725 Commonwealth Ave, Boston, MA 02215, USA}

\author{B. Decker}
\affil{Department of Physics and Astronomy, University of Missouri-Kansas City, 5110 Rockhill Road, Kansas City, MO 64110, USA}

\author{J.~D. Wing}
\affil{Harvard-Smithsonian Center for Astrophysics, 60 Garden Street, Cambridge, MA 02138, USA}

\author{G. Anand}
\affil{Astronomy Department and Institute for Astrophysical Research, Boston University, 725 Commonwealth Ave, Boston, MA 02215, USA}


\begin{abstract}
We present 190 galaxy cluster candidates (most at high redshift) based on galaxy {overdensity measurements} in the \spitzer/IRAC imaging of the fields surrounding 646 bent, double-lobed radio sources drawn from the Clusters Occupied by Bent Radio AGN (COBRA) Survey.  The COBRA sources were chosen as objects in the VLA FIRST survey that lack optical counterparts in the Sloan Digital Sky Survey (SDSS) to a limit of $m_r=22$, making them likely to lie at high redshift. This is confirmed by our observations: the redshift distribution of COBRA sources with estimated redshifts peaks near $z=1$, and extends out to $z\approx3$.   {Cluster candidates were identified by comparing our target fields to a background field and searching for statistically significant ($\ge2\sigma$) excesses in the galaxy number counts surrounding the radio sources; 190 fields satisfy the $\ge2\sigma$ limit.  We find that 530 fields (82.0\%) have a net positive excess of galaxies surrounding the radio source.  Many of the fields with positive excesses but below the $2\sigma$ cutoff are likely to be galaxy groups.}    Forty-one COBRA sources are quasars with known spectroscopic redshifts, which may be tracers of some of the most distant clusters known.

\end{abstract}

\keywords{galaxies:clusters:general -- galaxies:high-redshift -- galaxies:evolution -- infrared:galaxies -- radio continuum:galaxies}

\section{Introduction} \label{intro}

Galaxy clusters are the largest gravitationally bound structures in the universe.  They are composed of galaxies, hot X-ray emitting gas, and dark matter.  Clusters can be used to investigate many properties of our Universe, including the large-scale structure and dark matter distribution, galaxy formation and evolution, and cosmological parameters.  To explore these properties, it is useful to have a sample of clusters covering a large range of redshifts with a well-characterized selection technique.  Clusters are commonly detected via a variety of methods, each of which has its strengths and weaknesses.  These include {finding} galaxy overdensities in optical and infrared surveys {(the earliest of which were done by \citet{abell} and \citet{zwicky} in the optical)}, {detecting} extended emission in X-ray surveys~{\citep[e.g.,][]{vikhlinin2009}}, and {measuring} Sunyaev-Zel'dovich decrements in radio surveys~{\citep[e.g.,][]{vanderlinde, marriage, hasselfield, bleem}}.  Thousands of clusters have spectroscopic redshifts measured at $z<0.25$, and {an increasing number of large surveys are searching for clusters at high redshift (see many of the references mentioned in this section).}  However, the number of spectroscopically confirmed clusters with redshifts $z>1.0$ is still relatively small.

Clusters selected at {optical} wavelengths are identified on the basis of two-dimensional galaxy overdensities, but projection effects can cause overcounting. In addition to galaxy overdensities, at optical wavelengths the red sequence can be used to identify clusters~\citep{redseq}.  From the color of the red sequence, an approximate redshift of the cluster can be calculated.  At higher redshifts, however, it becomes difficult to identify cluster members, as they become too faint to detect.  Additionally, the peak of the galaxy spectral energy distribution shifts to the infrared, causing galaxies to drop out of optical surveys.  { It is also the case that at high redshift, clusters may not have prominent red sequences yet, as many galaxies may still be forming stars~\citep{mcgee, brodwin13, hennig}.}  

Infrared telescopes such as the \textit{Spitzer Space Telescope}~\citep{spitzer} can be used to find new clusters with $z>1$~\citep[e.g.,][]{stanford05, brodwin2006, muzzin09}, as the peak emission from galaxies gets redshifted into the infrared.  Infrared surveys select galaxies via their stellar mass.  Owing to a beneficial K-correction, galaxies observed in the 4.5~\micron band with \spitzer do not fade with redshift out to at least $z\sim1.4$~\citep[e.g.,][]{mancone}.  {This method is also independent of the star formation rate of the cluster galaxies (unless a secondary cut is applied, such as the optical/IR color cut in \citet{muzzin}), and thus will also find clusters whose red sequences are just starting to build up, or that do not have a red sequence, which is the case as one moves to higher redshift.}  \spitzer has allowed several large $z>1$ cluster surveys \citep[e.g.,][]{eisenhardt, muzzin, papovich, stanford, zeimann, rettura}.  While \spitzer enables very high-redshift searches over relatively small areas, the Wide-field Infrared Survey Explorer (WISE) is enabling an all-sky search for massive clusters at $z\sim1$~\citep[e.g.,][]{stanford14, brodwin15, gonzalez2015}; however, the spatial resolution of WISE is much poorer than that of \spitzer.

In the X-ray regime, galaxy clusters are the most common bright, extended extragalactic sources detected in surveys.  Clusters are identified through detection of the X-ray emitting gas in the intracluster medium (ICM).  The ICM is compressed and heated by the gravitational potential of the cluster and hence is a very clean cluster observable.  These extended sources are rare and so suffer much less from projection effects.  {While X-ray samples are complete at low redshifts~\citep{reiprich}}, X-ray observations of clusters require long exposure times, especially at high redshift, since the surface brightness declines as $(1+z)^4$.  Thus, flux-limited surveys are biased towards the most luminous (massive) clusters {(however, see \citealp{churazov}, who argue that the higher temperature and density of high-redshift clusters compensates for the surface brightness dimming)}.  {Additionally, they are also biased towards relaxed clusters, as those clusters that are not relaxed tend to have a less centrally peaked surface brightness profile.}

Both optical and X-ray surveys become more costly at high redshifts; however, the Sunyaev-Zel'dovich (SZ, \citealp{sz}) effect is nearly independent of cluster redshift and thus can probe the high-redshift regime.  The SZ effect is the distortion of the cosmic microwave background (CMB) spectrum as CMB photons are inverse Compton scattered as they pass through the ICM.  By looking for these distortions, new clusters can be found~\citep{bleem, hasselfield}.  The strength of the signal is proportional to the mass of the cluster, so lower-mass clusters may not be detected.  Additionally, bright radio point sources such as active galactic nuclei (AGN) can weaken the SZ signature for lower-mass clusters at high redshifts~\citep{linmohr}.  \citet{galametz09} and \citet{martini} showed that the AGN fraction of clusters increases with redshift, which would make this effect more common at higher redshift. 

Previous studies have found that radio sources, particularly bent, double-lobed radio sources, are frequently associated with galaxy clusters~\citep{blanton2000, blanton2001, blanton2003, giacintucci, wing, a3395, emu, dehghan}.  The radio lobes of these AGN are most likely bent because of the ram pressure that occurs due to the relative motion of the AGN host galaxy and the ICM, {which makes them good tracers for finding galaxy clusters}.  There are a few possible reasons for this relative motion.  First, it is possible that the lobes are bent by the ICM as the galaxy moves through it with a large peculiar velocity.  A second explanation is that the ICM is disrupted by a recent large-scale cluster-cluster merger \citep{burns}.  In this scenario, a galaxy with a low peculiar velocity encounters the large-scale bulk flow of the ICM, which is enough to bend the lobes.  Lastly, these bent, double-lobed radio sources can be found in clusters that are relatively relaxed on large scales, such as Abell 2029~\citep{clarke, rpm}.  In such clusters, the sloshing of the ICM (which is related to the merger history of the cluster, \citealp{am}) may cause the bending of the radio lobes.  {In addition to bent, double-lobed sources, powerful radio sources at high redshift that don't exhibit bending are also frequently associated with the cluster environment~\citep{venemans, w13, w14}.  It may be that the protocluster environment promotes high accretion rates at the scale necessary to generate radio-loud AGN~\citep{hatch}.  It is also possible that these objects are associated with protoclusters because they are among the most massive galaxies in the universe, and thus inhabit regions of the greatest overdensities, where cluster formation occurs~\citep{miley}.}  

Because of their frequent association with clusters and groups, {and the ease of detecting them in short exposures in large radio surveys,} bent, doubled-lobed radio sources make ideal tracers for finding high-redshift clusters.  Unlike many other cluster-finding methods, whose mass limit depends on the survey depth in the cluster mass observable, bent, double-lobed radio sources can be found in associations over a wide mass range, from lower-mass groups to the most massive clusters.  They are also found in a wide variety of environments, from clusters that are actively undergoing mergers~{ \citep[e.g.,][]{douglass}} to clusters that are highly relaxed on large scales~{ \citep[e.g.,][]{rpm}}.  Additionally, by the nature of their selection, the clusters in this sample will all contain radio-bright AGN, making them sources in which we can study AGN feedback.  Similar cluster searches have been done using powerful radio sources without regard to their morphologies~\citep{galametz, w13, w14}.

Here we present the initial results of the Clusters Occupied by Bent Radio AGN (COBRA) survey.  In \S\ref{sample} we present the sample.  In \S\ref{data}, we present our \spitzer data, and in \S\ref{cluster}, we discuss the number of candidate clusters found.  For cluster candidates observed in both the 3.6~\micron and 4.5~\micron bands we were able to make initial {photometric} redshift estimates, which we present in \S\ref{zest}.  In \S\ref{indiv} we present representative examples of our cluster candidates, and in \S\ref{conc} we present our conclusions.  Throughout, we assume a cosmology with $H_0=70$~km~s$^{-1}$~Mpc$^{-1}$, $\Omega_{\Lambda}=0.7$, and $\Omega_M = 0.3$.  All cosmological distances were calculated using the online Cosmological Calculator~\citep{cosmocalc}.  Magnitudes are given using the AB system and were calculated using SExtractor's automatic photometry routine\footnote{\textit{SExtractor} User's Manual, v2.13} (MAG\_AUTO).  The full COBRA catalog is available as an online supplement to this paper.

\section{The High-Redshift COBRA Sample} \label{sample}

Using the Faint Images of the Radio Sky at Twenty Centimeters (FIRST; \citealp{first}) survey, \citet{wing} created four samples of radio sources and examined their optical environments.  The FIRST survey covers $\sim25$\% of the sky and is mostly contained in the northern Galactic cap.  It has a flux density threshold of 1 mJy, systematic astrometric errors $<0.05''$, spatial resolution of 5\arcsec, and total positional errors on the order of $\sim1''$.  Each sample created by \citet{wing} consisted of a unique selection criterion:  visual-bent, auto-bent, straight, and single-component.  These samples were then cross-correlated with the Sloan Digital Sky Survey (SDSS) to find optical matches to a limit of $m_r=22$ to the presumed radio cores.  Sources associated with the bent radio sources without optical matches are likely to be distant and form the high-redshift COBRA sample. 

The visual-bent sample was compiled by visually examining a sample of $\sim32,000$ multiple-component radio sources from the 1997 April release of the FIRST catalog~\citep{lizthesis}.  Visual-bent sample sources were defined as those that have two or more radio components and a bent morphology.  From this, 384 sources were identified as bent, double-lobed sources. Of those,  272 had unique matches in the SDSS.  The 272 objects are included in the low-redshift COBRA sample~\citep{wing}; however in this work we consider only the 112 objects with $m_r$ fainter than 22, as they are likely to lie at high redshift.

The auto-bent sample was created by using a pattern recognition program~\citep{proctor} to identify bent, double-lobed sources over the entire FIRST catalog (as of 2003 April).  The sources in the auto-bent sample all contain three radio components (nominally a core and two lobes). The central component of these automatically detected sources is defined as the component opposite the longest side when making a triangle of the three components.  This sample contains 1546 sources of which 599 have unique SDSS matches.  {There are 94 sources in the auto-bent sample that overlap with the visual-bent sample.  For more detail, see \citet{wing}.}  As with the visual-bent sample, we only consider sources without SDSS matches as defined in \citet{wing}. {This set of sources was further trimmed to arrive at the sample of 653 targets (see below).}  In both the visual-bent and auto-bent samples the distance between any two components was limited to be no more than 60\arcsec.  

In \citet{wing}, the straight sample consists of all of the straight, three-component sources in the FIRST survey region.  The straight sources have an opening angle greater than 160$^\circ$, as compared to the bent sources, which have an opening angle less than 160$^\circ$. 

The single-component sample acts as a control.  If bent, double-lobed radio sources are preferentially associated with galaxy clusters, then the single-component sample should be associated with clusters at a lower rate than the bent sample.  The 782 sources in the single-component sample were randomly selected from the FIRST catalog and have no other radio source within $60''$.  They also have matches in the SDSS.  We do not consider either the straight sample or the single-component sample in this work.  

\citet{wing} cross-correlated their samples of radio sources with SDSS to examine their optical environments.  They found that the visual-bent sample was associated with clusters or groups with 20 or more member galaxies within a 1~Mpc radius of the radio source with an absolute \textit{r}-magnitude brighter than $M_r=-19$ 78\% of the time.  The association rate drops to 59\% for the auto-bent sample, 43\% for the straight sample, and 29\% for the single-component sample.  If we only include richer systems with 40 or more member galaxies the association rates are 62\% for the visual-bent sample, 41\% for the auto-bent sample, 24\% for the straight sample, and 10\% for the single-component sample. 

From the original \citet{wing} sample, {PI Blanton visually examined overlays of FIRST radio contours over SDSS images and eliminated the most obvious very low-redshift sources and sources that were not likely to be true bent, double-lobed radio sources. This selection process} identified 653 bent, double-lobed radio sources (including sources from the visual- and auto-bent samples) that did not have {obvious} detected optical galaxy hosts in the SDSS in the r-band to the limit of $m_r=22$ or were detected as blue quasar-like objects and have a known spectroscopic redshift $z>0.7$.  {For examples of the objects in the sample, see Figure~\ref{comparison} in \S\ref{indiv}.}  Most of the bent, double-lobed radio sources that had matches in SDSS consistent with elliptical galaxies found by \citet{wing} have redshift $z<0.7$.  {Since the detection limit of SDSS used in \citet{wing} is $m_r=22$, if any of the radio sources in the sample of 653 SDSS non-detections have counterparts that are consistent with elliptical galaxies, the elliptical galaxy will be fainter than $m_r=22$ in the SDSS and/or is associated with a different component of the radio emission than was assumed in \citet{wing}.}  Thus the radio sources without known optical identifications are most likely at high redshift.  There are some cases where the host galaxy was identified offset from the presumed radio core {using \spitzer data} and some of these hosts have $m_r<22$ as discussed in \S\ref{zest}.  The original cross-correlation in {\citet{wing} using SDSS Data Release 7~\citep{dr7}} yielded 32 blue, quasar-like objects {with spectroscopic redshifts in the SDSS} and 621 candidates without an optical host based on the earlier cross-correlation.  {Objects designated as quasars were designated as such because of their blue color and point-source appearance.}  Of these 653 sources, 646 were successfully observed in the infrared in our \spitzer snapshot program (\textsection{\ref{obs}}): 511 were observed only at 3.6~\micron, and 135 (including the 32 quasars) were observed at both 3.6~\micron and 4.5~\micron.  A detailed analysis {using visual identification of the radio host object and correlation with SDSS Data Release 12} yielded 9 more sources spectroscopically identified as quasars {(however, these objects were only observed at 3.6~\micron)}, for a total of 41 quasars and 605 radio AGN not identified as quasars.  Together, these 646 sources make up the high-redshift portion of the COBRA survey; the sources with SDSS matches analyzed in \citet{wing} make up the low-redshift portion.

\section{Data Reduction} \label{data}
\subsection{\textit{Spitzer} Observations} \label{obs}

The Infrared Array Camera (IRAC; \citealp{irac}) observations of the COBRA sources were carried out as a Snapshot Program during Cycle 8 (PID 80161, PI Blanton) using standard observing parameters.  Each of the fields targeted by our program was observed with multiple dithered full-array exposures; the number of exposures was adjusted for each source to achieve comparable on-source S/N ratios under different background conditions.  A medium dither throw was used.  All {646} sources were imaged at 3.6~\micron.  Two different exposure times were chosen for the sources observed at 3.6~\micron only: $5\times30$~s for sources estimated to have low background and $7\times30$~s for sources estimated to have a higher background.  We also observed 135 sources at 4.5~\micron.  The sources observed at both 3.6~\micron and 4.5~\micron were observed for $4\times100$~s {in both bands}.  {The exposure times for the sources observed at 3.6~\micron only were calculated with the goal of reaching a limiting 3.6~\micron flux density of 5~$\mu$Jy (M*$+$1.5) with a SNR = 5. The targeted limiting flux density for the fields observed for $4\times100$~s is 3.6~$\mu$Jy.  See \S\ref{bkgrnd} for a description of our adopted survey depth.}  These 135 sources include all the {sources originally identified as} quasars, sources for which we had existing complementary data (in either the optical or NIR), and sources that were deemed especially good candidates because of the appearance of their radio morphologies (particularly clear cases of bent, symmetric, double-lobed sources). In total, 646 distinct COBRA targets were observed between 2011 July and 2013 March. 

The IRAC imaging was reduced using standard techniques, using as a starting point the corrected basic calibrated data (cBCD). The cBCD frames were object-masked and median-stacked on a per-AOR (Astronomical Observing Request) basis; the resulting stacked images were then visually inspected and subtracted from individual cBCDs within each AOR. This was done to eliminate long-term residual images arising from prior observations of bright sources by the 3.6~\micron and 4.5~\micron arrays.  Subtracting the median stacks also minimized gradients in the celestial backgrounds around each source.  After these preliminaries, the data for each target were mosaicked into spatially registered mosaics using IRACproc~\citep{reduction}. IRACproc was configured to automatically flag and reject cosmic ray hits based on pipeline-generated masks together with a sigma-clipping algorithm for spatially coincident pixels. The cBCD frames were resampled to 0\farcs6 pixel$^{-1}$ during mosaicking, so each pixel in the final mosaic subtends one-fourth the solid angle of the native IRAC pixels.  The resulting mosaics and coverage maps were subsequently used as the basis for the photometric measurements described below.

\subsection{Source Extraction} \label{sextract}

We used SExtractor~\citep{sextractor} in single-image mode on the 3.6~\micron cutout frames for all observations.  We used frames trimmed to 4\farcm5 on a side.  We used many of the same parameters as outlined in \citet{lacy}, with a few changes.  We did apply a filter, using the \texttt{tophat\_2.0\_3x3.conv} file that is included with SExtractor.  The mosaics have a pixel scale of 0\farcs6 pixel$^{-1}$, so the tophat filter has a FWHM of 1\farcs2, which is smaller than the IRAC FWHM ($1\farcs95$).  This smooths the image before detection, making it easier to detect faint objects.  We also used a DEBLEND\_MINCONT value of 0.0001, a BACK\_SIZE value of 25 pixels, and applied the gain appropriate for our observations, based on the number of exposures and exposure time of each field.  The parameter DEBLEND\_MINCONT is the minimum contrast used when deblending pixels into different objects, and BACK\_SIZE is the size (in pixels) of the area used to estimate the background.  We ran SExtractor on the cutouts of our \spitzer images and created catalogs of the positions of each detected source, along with the fluxes and magnitudes of each detected source.  

When 4.5~\micron\ { observations were available}, we ran SExtractor in dual-image mode, using the 3.6~\micron image as the reference.  We used the same parameters described above.  {These catalogs were used for estimating redshifts (see \S\ref{zest}).}

\section{Cluster Candidates} \label{cluster}

We measured galaxy overdensities in the {3.6~\micron observations of the} COBRA fields by counting sources within 1\arcmin\ and 2\arcmin\ of our radio positions\footnote{At a redshift of $z=1.0$,  $1\arcmin = 480$~kpc.}, and comparing the results to mean galaxy counts at our search depth {(see \S\ref{bkgrnd})} in the deeper IRAC mosaics of the \textit{Spitzer} UKIDSS Ultra Deep Survey (SpUDS, PI: J. Dunlop).  COBRA fields with galaxy number counts greater than $2\sigma$ {(formally, $1.95\sigma$)} in excess of the background mean densities were identified as cluster candidates.  {In compact, rich clusters, the core radius is $R_c\simeq(0.1-0.3)h^{-1}$~Mpc~\citep{bahcall, dressler, sarazin}.  This corresponds to an angular extent of $\sim1\arcmin$ at $z=1.0$, which is the expected peak of the redshift distribution of our sample.  We extend the search to 2$\arcmin$ to account for the possibility that some of the AGN in our sample reside at the cluster outskirts, rather than at the center of the cluster.}

\subsection{Mean Background Counts} \label{bkgrnd}
To determine the mean background counts, we used the SpUDS field.  The SpUDS field covers the same approximately one square-degree patch of sky that the UKIDSS Ultra Deep Survey does, and consists of deep IRAC and 24~\micron MIPS observations.  We performed the same source extraction as described in \S\ref{sextract} on the final 3.6~\micron SpUDS mosaics to ensure consistency in our analyses. 

\begin{figure}
\begin{center}
\includegraphics[scale=0.6]{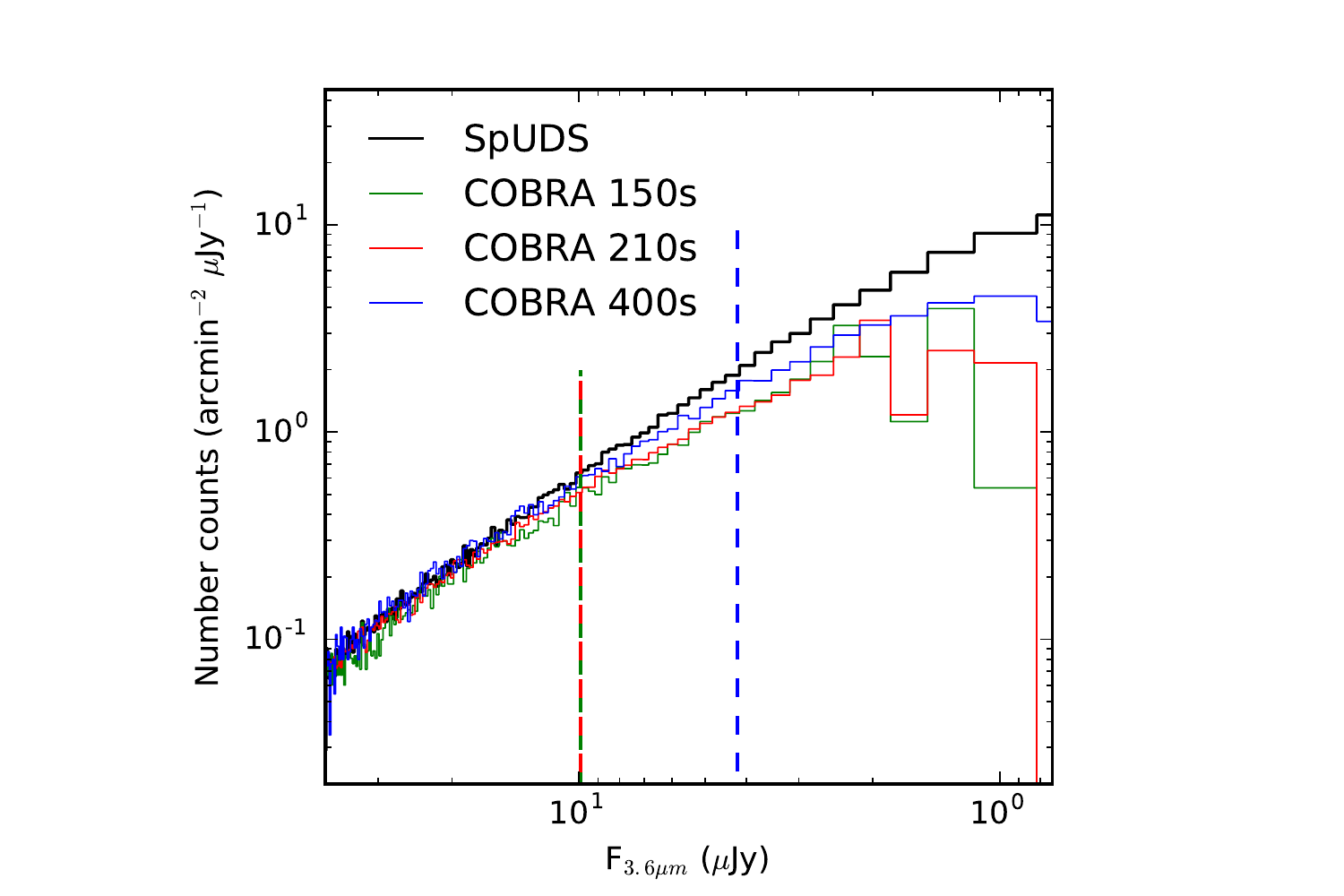}
\caption{{Number counts per square arcminute per $\mu$Jy for the different COBRA exposure times (green: $5\times30$~s, red: $7\times30$~s, blue: $4\times100$~s) and SpUDS (solid black) fields.  The vertical dashed lines show the flux at which COBRA number counts are 95\% those of the SpUDS survey.}}
\label{mag_comp}
\end{center}
\end{figure}

{The SpUDS field is much deeper than our COBRA fields, so we used it to determine our magnitude limit.  To do this, we compared the number of detected sources per square arcminute per $\mu$Jy bin for each of our \spitzer observation times and the SpUDS field, as shown in Figure~\ref{mag_comp}, {following the methodology of \citet{w13}, and described below}.  Figure~\ref{mag_comp} shows that at 9.6~$\mu$Jy, our $5\times30$~s COBRA fields, which are the shallowest in our survey, have 95\% of the number counts of the SpUDS field.  This corresponds to $m_{3.6}=21.4$, which we adopt as our limiting magnitude.  At this magnitude our fields with $7\times30$~s and $4\times100$~s exposure times have at least 95\% of the sources seen in the SpUDS field.}  

\begin{figure}
\begin{center}
\includegraphics[height=\linewidth, angle=0, trim={0.75in 0.25in 1in 0.25in}, clip=true]{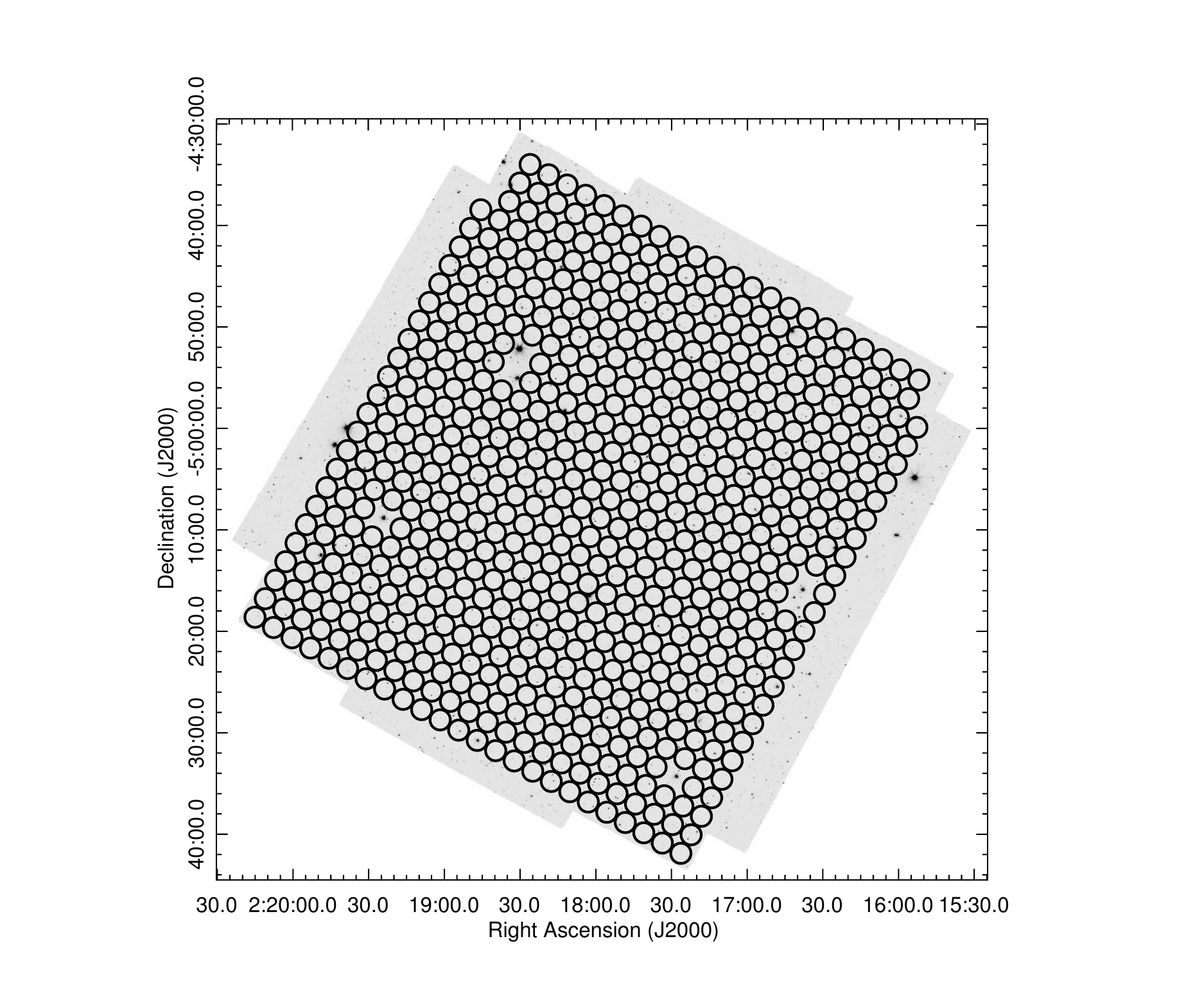}
\caption{The SpUDS field with 612 regions of 1\arcmin\ radius overlaid. Areas where bright foreground objects could possibly contaminate the source list were not included.}
\label{spuds}
\end{center}
\end{figure}

To measure the mean background surface count density, we counted up all of the sources in the SExtractor catalog within the 612 one arcminute regions that we placed on the SpUDS field, as shown in Figure~\ref{spuds}, to {$m_{3.6}=21.4$}.  Figure 17 in \citet{ashbymw} shows that at the depth of the COBRA survey, the number count for Milky Way stars is approximately 0.3~objects~mag$^{-1}$~arcmin$^{-2}$, which is a factor of approximately 100 below the number count we measure for the galaxies in the SpUDS field.  Thus no attempt was made to correct for the small contamination due to foreground stars.  We also avoided areas of the SpUDS field with bright foreground stars, as seen in Figure~\ref{spuds}.  Because the \spuds field is quite large, it is likely to contain groups and clusters.  To avoid having such structures affect our measurement of the mean density of the background field, we fitted a Gaussian to the lower half of the distribution of the SpUDS fields galaxy surface area densities (the dashed red histogram to the left of the vertical line in Figure~\ref{hist}; the solid red curve shows the Gaussian fit), following the method described in \citet{galametz}.  We find that the average background surface density is {8.5~sources~arcmin$^{-2}$ (26.7 sources per one arcminute radius region) with a standard deviation {($\sigma_{SpUDS}$)} of 2.0~sources~arcmin$^{-2}$ (6.3 sources per one arcminute radius region).  We performed a similar analysis on the SpUDS fields with regions two arcminutes in radius.  Using 152 two arcminute regions we find a mean background surface density of 8.4~sources~arcmin$^{-2}$ (105.6 sources per two arcminute radius region) and a standard deviation {($\sigma_{SpUDS}$)} of 1.2~sources~arcmin$^{-2}$ (15.1 sources per two arcminute radius region).}
  
\subsection{Identifying Cluster Candidates} \label{clustercan}

\begin{figure}
\begin{center}
\includegraphics[height=\linewidth, angle=0, trim={0.15in 0.25in 0.23in 0.25in}, clip=true]{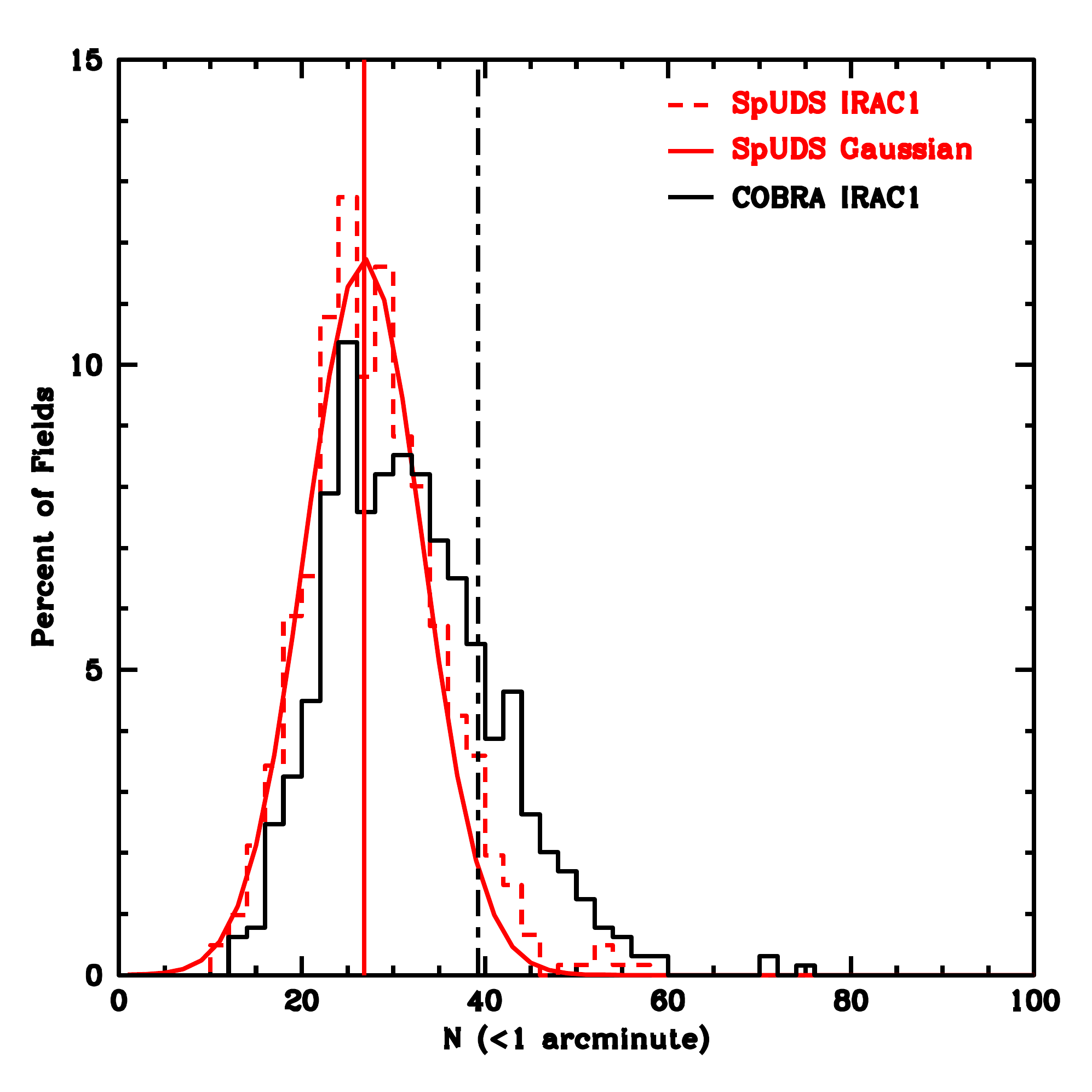}
\caption{Histogram of the percentage of fields with a given number of detected sources in a one arcminute radius region for a limiting magnitude of $m_{3.6}=21.4$.  The solid black histogram shows results from the 646 COBRA fields, while the dashed red histogram shows results from the 612 regions in the SpUDS field.  The solid, red curve shows the Gaussian fit described in \S\ref{bkgrnd}.  The solid, red vertical line shows the mean of the Gaussian fit, and the {dot-dash, black} vertical line marks an excess of {12.4} sources, which corresponds to an overdensity of 2$\sigma$. {Objects to the right of this line are cluster candidates.}}
\label{hist}
\end{center}
\end{figure}

\begin{figure}
\begin{center}
\includegraphics[height=\linewidth, angle=0, trim={0.15in 0.25in 0.23in 0.25in}, clip=true]{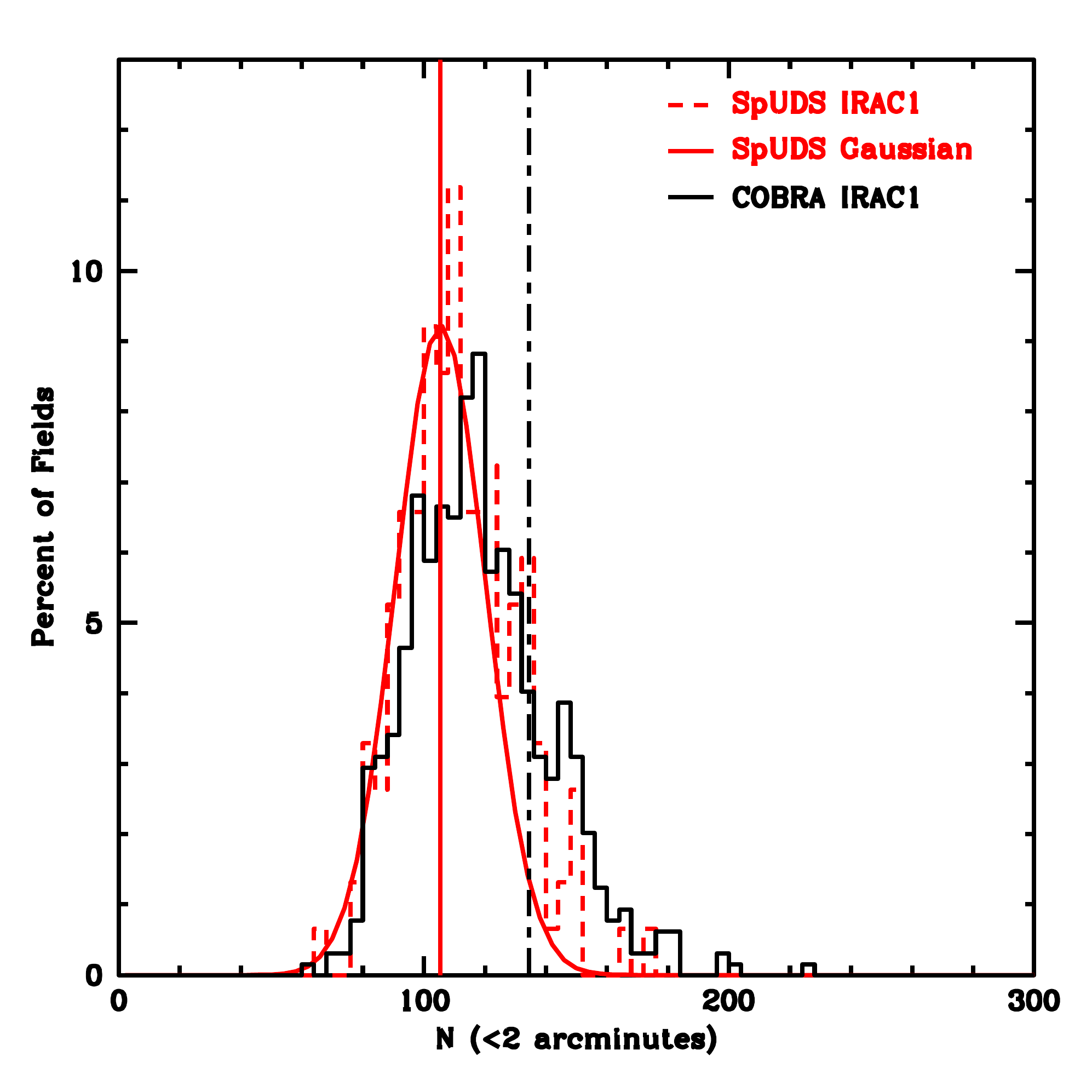}
\caption{Histogram of the percentage of fields with a given number of detected sources in a two arcminute radius region to a limiting magnitude of $m_{3.6}=21.4$.  The solid black histogram shows results from the 646 COBRA fields, while the dashed red histogram shows results from the 152 regions in the SpUDS field.  The solid, red curve shows the Gaussian fit described in \S\ref{bkgrnd}.  The solid, red vertical line shows the mean of the Gaussian fit, and the {dot-dash, black} vertical line marks an excess of {29.4} sources, which corresponds to an overdensity of 2$\sigma$.  {Objects to the right of this line are cluster candidates.}}
\label{hist2}
\end{center}
\end{figure}

Candidate clusters were identified on the basis of excess galaxy counts measured in the IRAC mosaics.  Specifically, we counted IRAC-detected sources within radii of both 1\arcmin~and 2\arcmin~of each COBRA radio source and compared these counts to the background counts from SpUDS.  As discussed in \citet{wing2013}, the radio sources in the low-redshift COBRA sample are located at both the center and the outskirts of the clusters, which is why we examine both one and two arcminute radius regions, and why some fields may have an overdensity within two arcminutes but not within one arcminute.  Figures~\ref{hist} and \ref{hist2} compare the source count distributions over all the COBRA fields to those measured with an identical procedure but in the SpUDS mosaics.  
Candidate galaxy clusters were selected as those having either an overdensity of $\ge2\sigma$ within 1\arcmin\ (an excess of {12.4} sources) or an overdensity of $\ge2\sigma$ within 2\arcmin ({29.4} sources), or an excess of sources within both radii, to a limiting magnitude of $m_{3.6}=21.4$. 

We calculated the error and significance of our overdensities as follows.  First, to determine the error on the overdensity, we assumed Poisson statistics, including the counts in the field surrounding each radio source and the background counts as determined from the SpUDS field, and {calculated the error on $\Delta N$, $\sigma_{\Delta N}$ (columns 10 and 13 in Table~\ref{overdensity}), with the following equation}:
\begin{equation}
\sigma_{\Delta N} = \sqrt{\Delta N + A\left(1+A/A_{bkg}\right)\Sigma_{avg}}.
\label{poisson}
\end{equation}
In Equation~\ref{poisson}, $\Delta N$ is the number of detected sources above the mean background, $A$ is the area of the COBRA field in which sources were counted, $A_{bkg}$ is the area of the entire SpUDS field within which sources were counted, and $\Sigma_{avg}$ is the average surface density of the SpUDS field.
Second, we use the sources detected around each radio source and the one-sigma variance of the Gaussian fit to determine the significance (columns 10 and 13 in Table~\ref{overdensity}) using the following equation:
\begin{equation}
\rm{Significance} = \frac{\Delta N}{\sigma_{SpUDS}}.
\label{gauss}
\end{equation}
This is the number of detected sources above the mean background divided by the standard deviation of the Gaussian fit.

Table~\ref{overdensity} shows the results of our search.  Column 1 lists the name of the source, while Columns 2 and 3 list the coordinates of the optical/IR host (see \S\ref{zest} for a detailed discussion of how the optical host coordinates were obtained).  Column 4 lists the 20~cm flux density as measured by FIRST and Column 5 lists the magnitude of the host in the 3.6~\micron band as determined by SExtractor.  Where available, host color {($m_{3.6} - m_{4.5}$)} and redshift are listed in columns 6 and 7.  Columns $8-10$ give the results from the search within a one arcminute radius, while columns $11-13$ give the results from the search within a two arcminute radius to a limiting magnitude of $m_{3.6}=21.4$.

\begin{longrotatetable}
\tabletypesize{\normalsize}
\rotate
\begin{deluxetable*}{l@{\extracolsep{2pt}}ccrrccrrrrrr}
\setlength{\tabcolsep}{0.05in}
\tablecolumns{13}
\tablecaption{COBRA Cluster Survey Catalog\label{overdensity}} 
\tablewidth{0pt}
\tablehead{
	\colhead{} & 
	\multicolumn{2}{c}{Host Coordinates}  &
	\colhead{} &
	\colhead{} &
	\colhead{} &
	\colhead{} &
	\multicolumn{3}{c}{One Arcmin} & 
	\multicolumn{3}{c}{Two Arcmin} \rule[-1.2ex]{0pt}{0pt}
	\cr
	\cline{2-3} \cline{8-10} \cline{11-13}\\
	\colhead{Object} & 
	\colhead{RA (J2000)} &
	\colhead{DEC (J2000)} &
	\colhead{$S_{20}$ (mJy)} &
	\colhead{$m_{3.6}$}&
	\colhead{$m_{3.6}-m_{4.5}$} & 
	\colhead{z\tablenotemark{a}} & 
	\colhead{$\Delta$N} & 
	\colhead{$\sigma_{\Delta N}$} &
	\colhead{Sig.} & 
	\colhead{$\Delta$N} & 
	\colhead{$\sigma_{\Delta N}$} &
	\colhead{Sig.} \rule{0pt}{2.6ex}
	\cr
	\colhead{(1)} &
	\colhead{(2)} &
	\colhead{(3)}&
	\colhead{(4)}&
	\colhead{(5)}&
	\colhead{(6)}&
	\colhead{(7)}&
	\colhead{(8)}&
	\colhead{(9)}&
	\colhead{(10)}&
	\colhead{(11)}&
	\colhead{(12)}&
	\colhead{(13)}
}
\startdata
\cutinhead{3.6~\micron Only}
COBRA002211.9$-$095013&00 22 09.89&$-$09 50 03.5&  54.8&17.65&\nodata& 0.43& $-$3.8& 4.8&$-$0.60& $-$8.5&10.0&$-$0.57\\
COBRA003225.8$-$110852&00 32 25.80&$-$11 08 52.4&  55.8&18.83&\nodata&\nodata& $-$1.8& 5.0&$-$0.28&$-$25.5& 9.1&$-$1.73\\
COBRA012058.9$+$002140&01 20 58.87&$+$00 21 41.7&  43.9&17.96&\nodata&  0.80& 27.2& 7.4& 4.27& 29.5&11.7& 2.00\\
COBRA014147.2$-$092812&\nodata\tablenotemark{b}&\nodata\tablenotemark{b}&   4.6&\nodata&\nodata&\nodata&  5.2& 5.7& 0.82& 16.5&11.2& 1.12\\
COBRA014339.2$-$011749&01 43 24.21&$-$01 18 15.8&  59.5&19.99&\nodata&\nodata& $-$5.8& 4.6&$-$0.91&$-$23.5& 9.2&$-$1.59\\
COBRA014741.6$-$004706&01 47 41.73&$+$00 47 08.5&  25.9&22.92&\nodata&\nodata& 10.2& 6.1& 1.60& 14.5&11.1& 0.98\\
\cutinhead{3.6~\micron and 4.5~\micron}
COBRA003447.7$-$002137&00 34 48.07&$+$00 21 32.0&  36.8&19.33& 0.15& 1.86\tablenotemark{d}& $-$1.8& 5.0&$-$0.28& $-$9.5& 9.9&$-$0.64\\
COBRA003625.8$-$005226&00 36 25.37&$+$00 52 30.7&  15.5&19.74&\nodata&\nodata&  7.2& 5.8& 1.13& 20.5&11.3& 1.39\\
COBRA005653.3$-$010455&00 56 53.28&$-$01 05 05.3&  12.9&19.34&\nodata&\nodata& $-$3.8& 4.8&$-$0.60& 17.5&11.2& 1.19\\
COBRA005837.2$+$011326&00 58 37.03&$+$01 13 27.8&  32.6&18.02&$-$0.54& 0.71& 14.2& 6.4& 2.23& 39.5&12.2& 2.68\\
COBRA015238.3$+$002617&01 52 38.26&$+$00 26 33.0&  73.2&18.64&$-$0.02& 1.37\tablenotemark{d}& 14.2& 6.4& 2.23& 19.5&11.3& 1.32\\
COBRA025356.3$-$011350&02 53 56.30&$-$01 13 27.8& 130.7&21.70& 0.16&  1.90\tablenotemark{d}& $-$8.8& 4.2&$-$1.38&$-$17.5& 9.5&$-$1.18\\
\cutinhead{Quasars}
COBRA010329.5$+$004055&01 03 29.50&$+$00 40 55.0& 115.3&18.21& 0.41&1.433&  0.2& 5.2& 0.03&  5.5&10.7& 0.38\\
COBRA012248.0$-$093546&01 22 48.21&$-$09 35 46.9&  37.0&17.06& 0.11&0.784&  4.2& 5.6& 0.66& $-$9.5& 9.9&$-$0.64\\
COBRA073320.4$+$272103&07 33 20.49&$+$27 21 03.6& 275.2&19.21& 0.09&2.938& 23.2& 7.1& 3.64& 56.5&12.8& 3.83\\
COBRA075228.6$+$375053&07 52 28.60&$+$37 50 53.0& 395.7&17.11& 0.32&1.208&  3.2& 5.5& 0.50& 44.5&12.4& 3.02\\
COBRA082643.4$+$143427&08 26 43.46&$+$14 34 27.6& 113.4&17.55& 0.13&2.312& $-$0.8& 5.1&$-$0.12& 18.5&11.3& 1.25\\
COBRA083951.6$+$292818&08 39 51.60&$+$29 28 18.0& 138.7&20.93& 0.36&1.136& $-$1.8& 5.0&$-$0.28&  1.5&10.5& 0.10\\
COBRA090102.7$+$420746&09 01 02.70&$+$42 07 46.0&  20.6&19.26& 0.37&1.621& $-$1.8& 5.0&$-$0.28& 30.5&11.8& 2.07\\
COBRA090745.5$+$382740&09 07 45.50&$+$38 27 40.0& 160.6&17.75& 0.29&1.743&  6.2& 5.7& 0.97& 33.5&11.9& 2.27\\
COBRA093023.2$+$484723&09 30 23.23&$+$48 47 23.6&  24.6&19.67&\nodata&  1.70\tablenotemark{k}& $-$3.8& 4.8&$-$0.60&$-$23.5& 9.2&$-$1.59\\
\enddata
\tablecomments{Column 1 identifies the radio source.  Columns 2 and 3 provide the coordinates of the optical/{IR} host, which in some cases differs from that of the radio source (see text).  Column 4 lists the 20~cm flux density as measured by FIRST and Column 5 lists the magnitude of the host in the 3.6~\micron band as determined by SExtractor.  Columns 6 and 7 list the host color and redshift where available.  Columns 8-10 give the results from the search within a 1\arcmin\ radius and columns 11-13 give the results for a 2\arcmin\ radius to a limiting magnitude of $m_{3.6}=21.4$.  Columns 9 and 12 list the total error on the overdensity, including Poisson noise contributions from both the cluster and background subtraction.  The significance in columns 10 and 13 were computed using Equation~\ref{gauss}, which relates the overdensity, $\Delta N$, to the standard deviation of the SpUDS field in each aperture size.
\\
\\
Table 1 is published in its entirety in machine-readable format online. A portion is shown here for guidance regarding its form and content.}

 \tablenotetext{a}{Photometric redshift available in SDSS, unless otherwise noted.  Quasar redshifts are spectroscopically determined.}
 \tablenotetext{b}{No host could be identified.}
 \tablenotetext{c}{Spectroscopic redshift from SDSS.}
 \tablenotetext{d}{Redshift is derived from the color of the source associated with the radio contours.  See text for more details.}
\tablenotetext{e}{While the listed color is too red to determine a redshift based on our models, examining the surrounding sources yields the listed redshift.}
\tablenotetext{f}{Identified in the SDSS Digital Release 6 Galaxy Clusters Catalog.}
\tablenotetext{g}{The listed coordinates of the radio source are not those of the identified host.  The identified host of the radio source was not at the center of the search radius used to calculate an overdensity.}
\tablenotetext{h}{Color could not be determined due to nearby saturated pixels.}
\tablenotetext{i}{Object identified as host is identified as a star in SDSS.}
\tablenotetext{j}{Spectroscopically determined redshift from \citet{blanton2003}. Redshift from the color is 0.99.}
\tablenotetext{k}{Source observed at 3.6~\micron only later identified as a quasar.}
\end{deluxetable*}
\end{longrotatetable}

To a limiting magnitude of $m_{3.6}=21.4$ we identify 48 fields that only satisfy the 1\arcmin~overdensity criterion, 47 that only satisfy the 2\arcmin~criterion, and 95 which met both criteria.  In total, 190 fields meet at least one of the two overdensity criteria with a limiting magnitude of $m_{3.6}=21.4$, {giving a total cluster association rate of 29.4\% to our adopted limiting magnitude}. 

\begin{deluxetable}{lccccc}
\tablecolumns{6}
\tablecaption{Number of Cluster Candidates}
\tablehead{
	\colhead{Sample} &
	\colhead{$1\arcmin$ Only} &
	\colhead{$2\arcmin$ Only} &
	\colhead{Both} &
	\colhead{Total} &
	\colhead{\%}
}
\startdata
All COBRA & 48 & 47 & 95 & 190 & 29.4\\
Non-quasars & 48 & 39 & 90 & 177 & 29.3 \\
Quasars & 0 & 8 & 5 & 13 &31.7\\
\enddata
\tablecomments{Column 1 identifies the subsample.  Columns 2- 4 provide the number of cluster candidates within the different search radii. Column 5 gives the total number of cluster candidates for each subsample. Column 6 is the percentage of the targets in the subsample that are cluster candidates.}
\label{percentage} 
\end{deluxetable}

\begin{figure}
\begin{center}
\includegraphics[height=\linewidth, angle=0, trim={0.05in 0.25in 0.23in 0.25in}, clip=true]{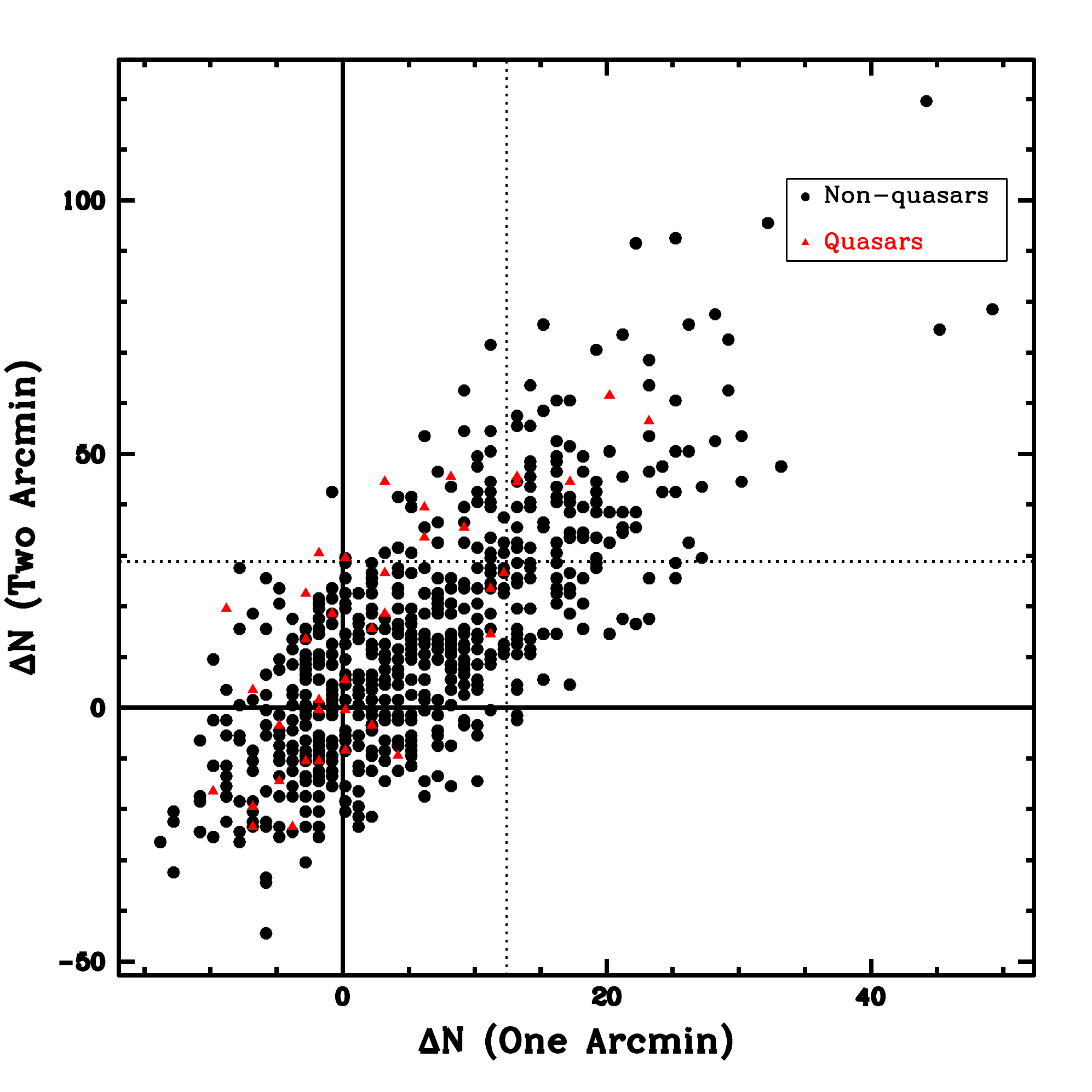}
\caption{Distribution of excess sources above background within two arcmin plotted against the excess above background within one arcmin in the COBRA fields to a limiting magnitude of $m_{3.6}=21.4$.  {The overdensities of fields surrounding quasars are} plotted as red triangles, while {the overdensities of fields around non-quasars} are plotted as black circles.  The vertical dashed line marks an overdensity of 2$\sigma$ within one arcmin (i.e. anything to the right would be considered a cluster searching within a one arcmin radius region) and the horizontal dashed line marks an overdensity of 2$\sigma$ within two arcmin (i.e. anything above that line would be a considered a cluster with a search radius of two arcmin). $\Delta N=0$ indicates that the number of sources surrounding the AGN is equal to the mean background value.  Any $\Delta N > 0$ indicates an excess above the mean background (solid black lines).  {Due to the scaling of the axes, some of the points for the quasars overlap with each other.}}
\label{clusters}
\end{center}
\end{figure}

Figure~\ref{clusters} shows the distribution of the overdensities surrounding each bent AGN to a limiting magnitude of $m_{3.6}=21.4$.  {The majority of the cluster candidates show overdensities within both search radii (defined as $\Delta N \ge 12.4$ within one arcminute and $\Delta N \ge 29.4$ within two arcminutes).  This is also true of the quasars.}  Excluding all the sources identified as quasars, there are 48 radio sources that are surrounded by overdensities within one arcminute only, 39 radio sources that are surrounded by overdensities within two arcminutes only, and 90 radio sources that are surrounded by overdensities within both search radii.  

Of the 41 quasars that meet our selection criteria, 13 are cluster candidates, for an association rate of 31.7\%.  This is comparable to the association rate of the nonquasars {(29.3\%)}.  The five that are cluster candidates within one arcminute also satisfy the candidate selection criterion within two arcminutes.  There are an additional eight cluster candidates that are not overdense within one arcminute but are within two arcminutes. This is tentative evidence that quasars tend to reside on the outskirts of clusters, which is to be expected if they are triggered during infall as clusters build up. Further observations will be helpful in testing this hypothesis, {which is very preliminary. Golden-Marx et al. (in preparation) will explore the relationship between the the radio source position and the projected surface density of sources and will compare the quasar/non-quasar differences in more detail.}  {Table~\ref{percentage} summarizes the number of cluster candidates found in each subsample.} 

There are 376 COBRA fields (including the quasars) with $\Delta N \ge 0$ within both search radii, 77 with $\Delta N \ge 0$ within one arcminute but with underdensities ($\Delta N < 0$) within two arcminutes, and 77 with $\Delta N \ge 0$ within two arcminutes but underdensities within one arcminute, {for a total of 530 fields (82.0\%) that have an excess number of sources as compared to the background.} There are 4 quasars that have $\Delta N \ge 0$ within one arcminute {but are underdense within two arcminutes}, and 9 more that have $\Delta N \ge 0$ within two arcminutes but are underdense within one arcminute.  There are an additional 18 quasars that have $\Delta N \ge 0$ within both search radii.  {These results are summarized in Table~\ref{od}.}  It is likely that the targets that only have overdensities in the larger search radius are not as centrally peaked, and are in the process of merging.  {Alternatively, the bent radio source may just be located away from the center of a (possibly relaxed) cluster.}

\begin{deluxetable}{lccccc}
\tablecolumns{6}
\tablecaption{Number of Targets with Positive Overdensities ($\Delta N \ge 0$)}
\tablehead{
	\colhead{Sample} &
	\colhead{$1\arcmin$ Only} &
	\colhead{$2\arcmin$ Only} &
	\colhead{Both} &
	\colhead{Total} &
	\colhead{\%}
}
\startdata
All COBRA & 77 & 77 & 376 & 530 & 82.0\\
Non-Quasars & 73 & 68 & 358 & 499 & 82.5\\
Quasars & 4 & 9 & 18 & 31 & 75.6\\
\enddata
\tablecomments{Column 1 identifies the subsample.  Columns 2-4 provide the number of targets with positive overdensities within the different search radii. Column 5 gives the total number of targets with positive overdensities for each subsample. Column 6 is the percentage of targets in that subsample with positive overdensities.}
\label{od} 
\end{deluxetable}

\section{Redshift Estimates}
\label{zest}

\begin{figure}
\begin{center}
\includegraphics[height=\linewidth, angle=0, trim={0.0in 0.25in 0.25in 0.25in}, clip=true]{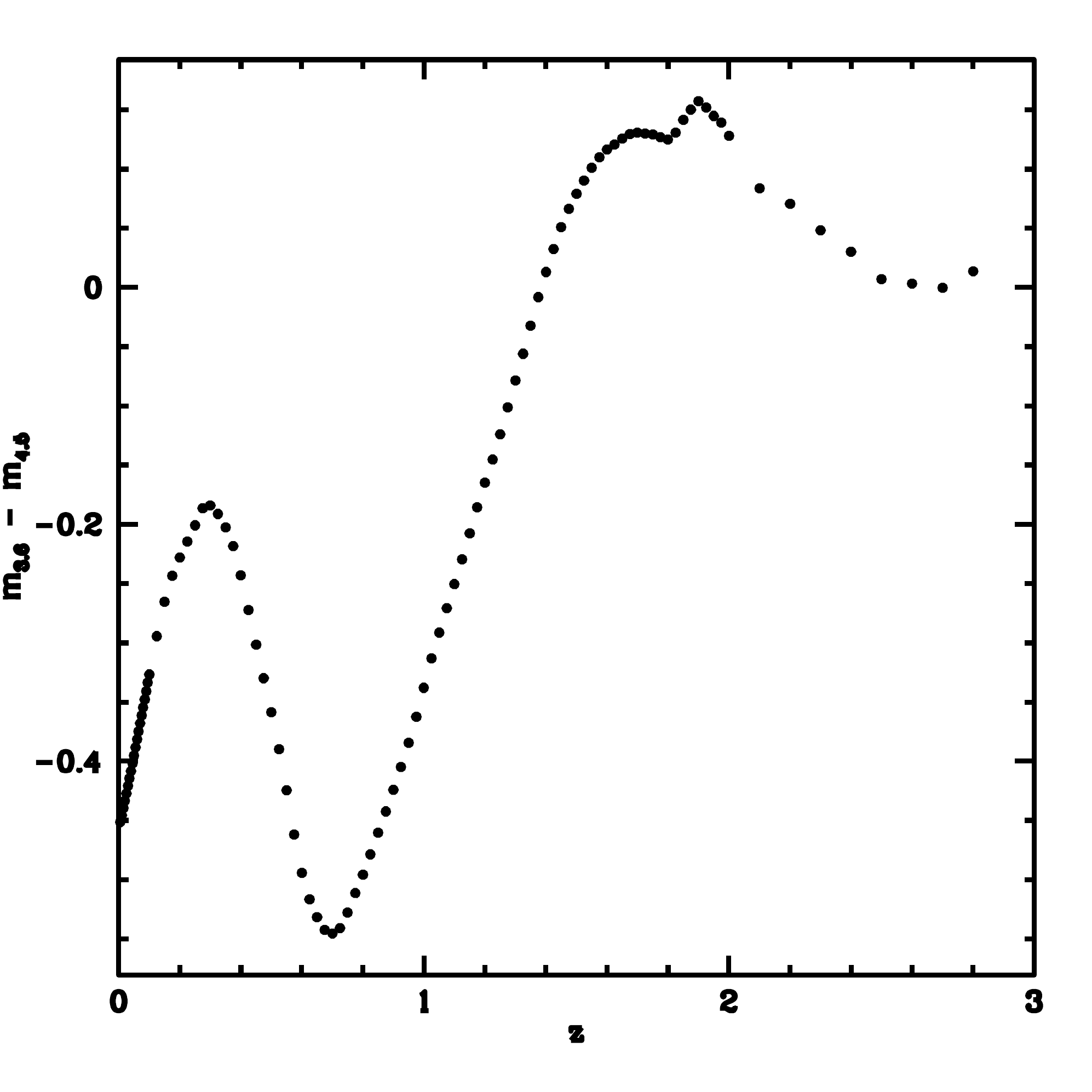}
\caption{Color {($m_{3.6} - m_{4.5}$)} as a function of redshift for a galaxy model with a Salpeter initial mass function, metallicity of 0.02, and a \citet{bc} simple stellar population model with a formation redshift of $z=3$.}
\label{colorz}
\end{center}
\end{figure}

For the COBRA targets that were observed in both the 3.6~\micron and 4.5~\micron bands, we used the color of the radio host to estimate a preliminary redshift.  To find the host galaxy, we performed a radial search (expanding in radius until a match was returned) around the coordinates of the radio source as defined in \S\ref{sample}, assuming that the closest match was the host.  We then visually inspected the radio contours overlaid on the \spitzer images to determine if there was a better match.  If there was a better match, its coordinates were obtained by searching the SExtractor catalog for the nearest SExtractor source detected within 2\arcsec\ of the coordinates of the new host.  The SDSS was searched again for redshifts after all hosts were identified.  Updated host coordinates are listed in Columns 2 and 3 of Table~\ref{overdensity}.  All of the 32 original quasars have optical hosts at the same coordinates (to within 2\arcsec) as the radio coordinates.   There are 15 non-quasars whose optical/infrared host could not be identified.  Once the host was determined, we used the galaxy model code \texttt{EzGal}~\citep{ezgal} with a Salpeter initial mass function, metallicity of 0.02, and a \citet{bc} simple stellar population model with a formation redshift of $z=3$ to relate our measured colors to redshifts.  Based on the limiting magnitude of SDSS, we assume that all host galaxies have a redshift $z>0.7$, unless otherwise noted.  At $z\gtrsim1.4$ the relation between color and redshift is not monotonic (Figure~\ref{colorz}), and occasionally there were two solutions for a given color.  For those hosts we list only the lower redshift.

Some sources in Table~\ref{overdensity} have either spectroscopic or photometric redshifts below our original redshift cutoff estimate of $z\approx0.7$.  These sources were not found in the original search by \citet{wing} due to either large offsets in the position of the estimate of the radio source core or the magnitude limit cutoff of $m_r=22$ in the original search.  Here we do not impose a magnitude limit in the SDSS.  Additionally, while the radio galaxy hosts are often giant ellipticals, they can have a wide range of luminosities and thus it is possible some have magnitudes fainter than $m_r=22$ and redshifts $z<0.7$.  The core of the radio source was determined visually in the visual-bent sample; however in the auto-bent sample the coordinates of the core were determined to be the component opposite the longest side when making a triangle of the three components.  After inspection of the \spitzer images, the host galaxy was sometimes found to be associated with another radio component.  

The majority of the redshifts determined from the infrared color of the source are consistent with the redshift listed in SDSS (within errors, { which are $\sim$10\%}).  For the few that are not, we keep the SDSS redshift, as they have more filters and thus a more accurate photometric redshift.

\begin{figure}
\begin{center}
\includegraphics[height=\linewidth, angle=0, trim={0.2in 0.25in 0.25in 0.25in}, clip=true]{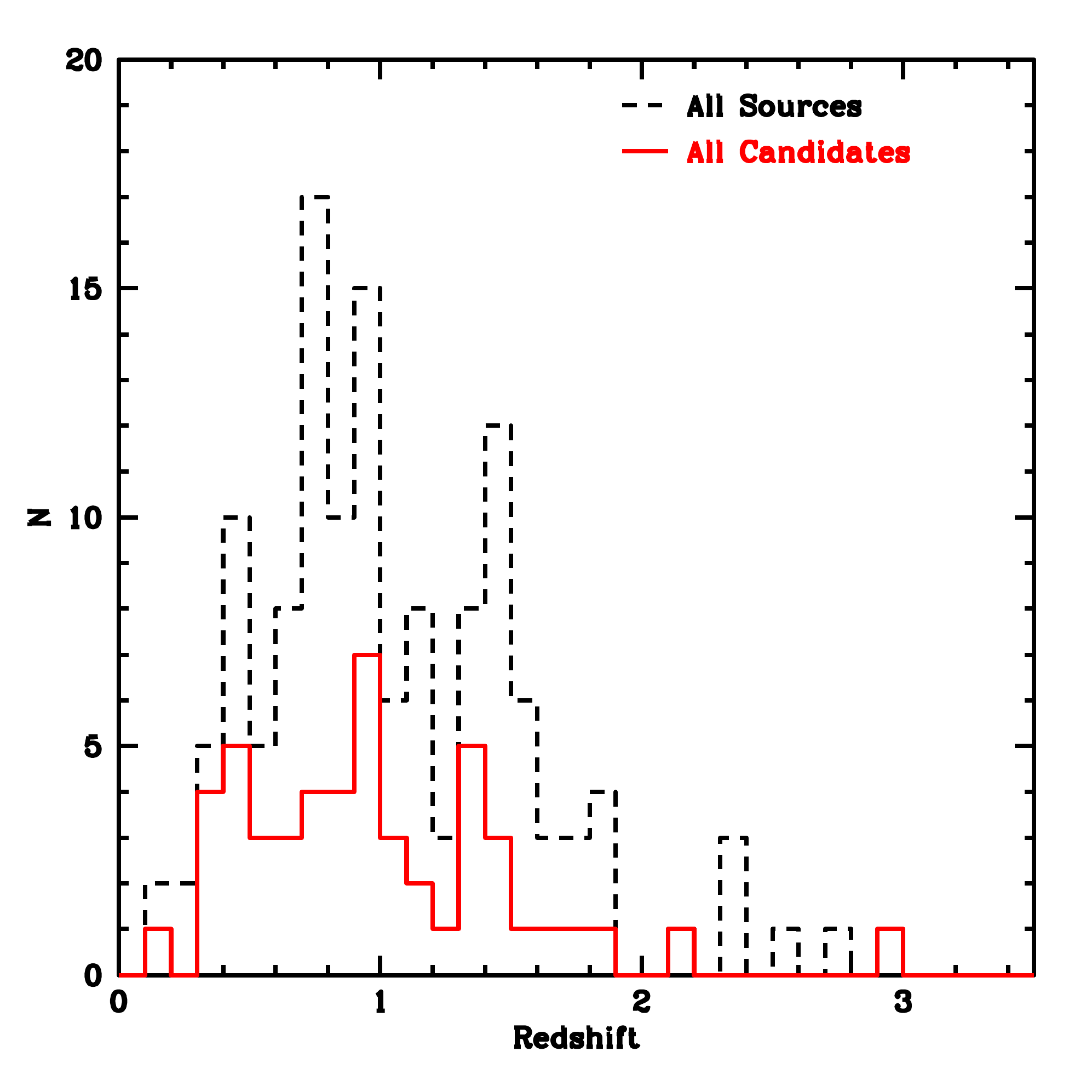}
\caption{Histogram of the redshift distribution of all sources that were observed in both the 3.6~\micron and 4.5~\micron bands with an estimated (or known, in the case of the quasars and sources found in the SDSS) redshift.  The dashed black line includes all sources; the solid red line shows the distribution of sources identified as cluster candidates.  For sources with degenerate redshift estimates, we used the lower one for the histogram.  Our objects extend in redshift up to $z\approx3.0$.}
\label{zhist}
\end{center}
\end{figure}

Figure~\ref{zhist} shows the redshift distribution of all IRAC-detected radio hosts in the COBRA fields that were observed in both bands (3.6~\micron and 4.5~\micron) for which we were able to estimate redshifts.  We include all the sources (dashed black line); we also separately show the distribution for our candidate clusters (solid red line).  All of the quasars in our sample have redshifts greater than $z=0.7$; 13 of the regions surrounding the quasars were identified as cluster candidates.  There are 103 non-quasar sources observed in both bands.  Of these, 41 were identified as cluster candidates.  We were only able to estimate redshifts for 94 of the {103 non-quasar} sources; the remaining sources were either too red to use the normal galaxy color-redshift relation or did not have an identifiable host.  Three of these nine are cluster candidates.  Of the 38 {non-quasar} cluster candidates with redshifts, 22 have $z>0.7$, 7 have $0.5\leq z < 0.7$, and {9 have $z<0.5$}.  All of the ones with redshifts $z<0.7$ were either spectroscopically or photometrically identified in the SDSS. 

\section{Individual Clusters}
\label{indiv}

\begin{figure*}
\begin{center}
\subfigure{\includegraphics[scale=0.4, trim={0.0in 0.25in 0.0in 0.25in}, clip=true]{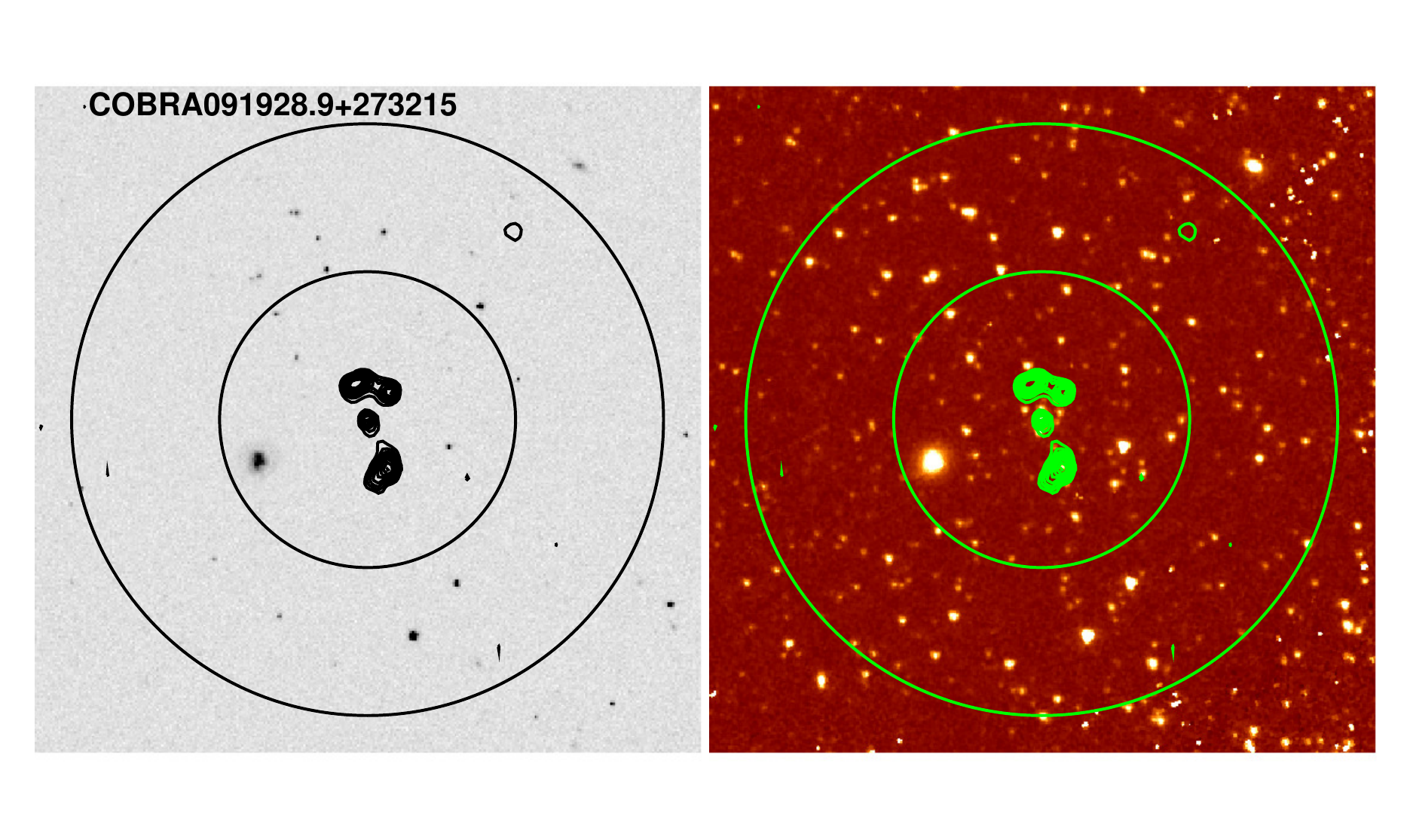}}
\subfigure{\includegraphics[scale=0.4, trim={0.0in 0.25in 0.0in 0.25in}, clip=true]{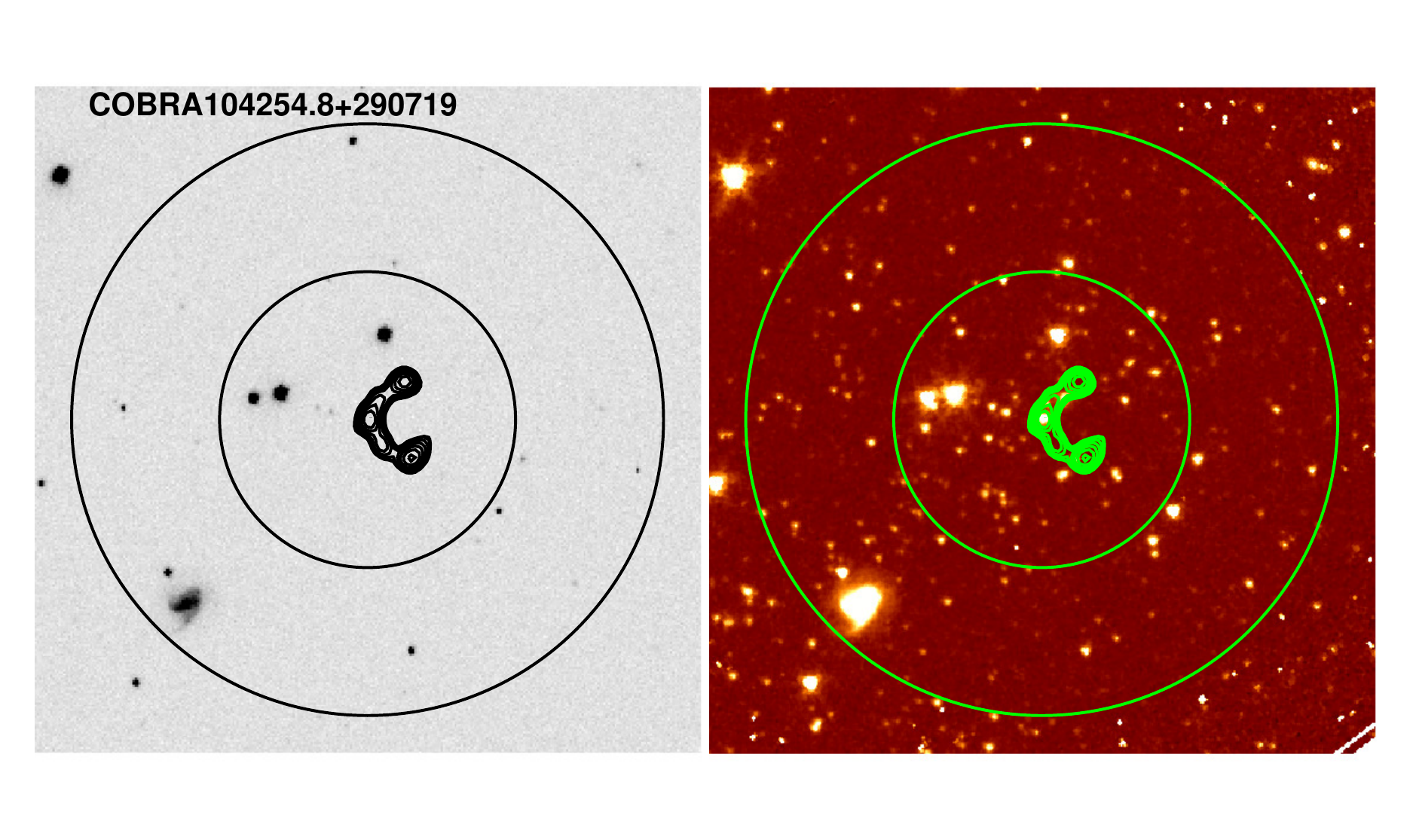}}
\subfigure{\includegraphics[scale=0.4, trim={0.0in 0.25in 0.0in 0.25in}, clip=true]{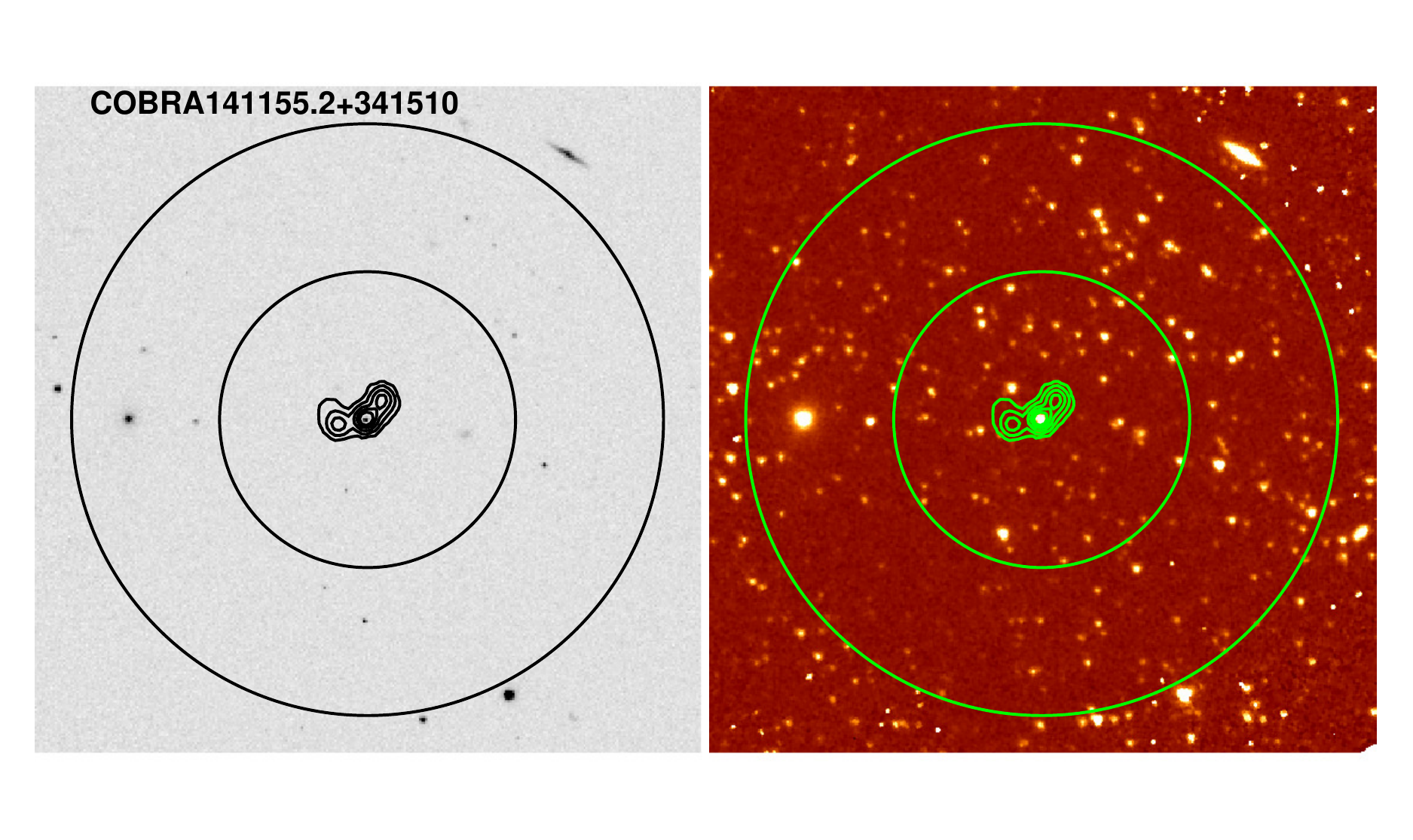}}
\subfigure{\includegraphics[scale=0.4, trim={0.0in 0.25in 0.0in 0.25in}, clip=true]{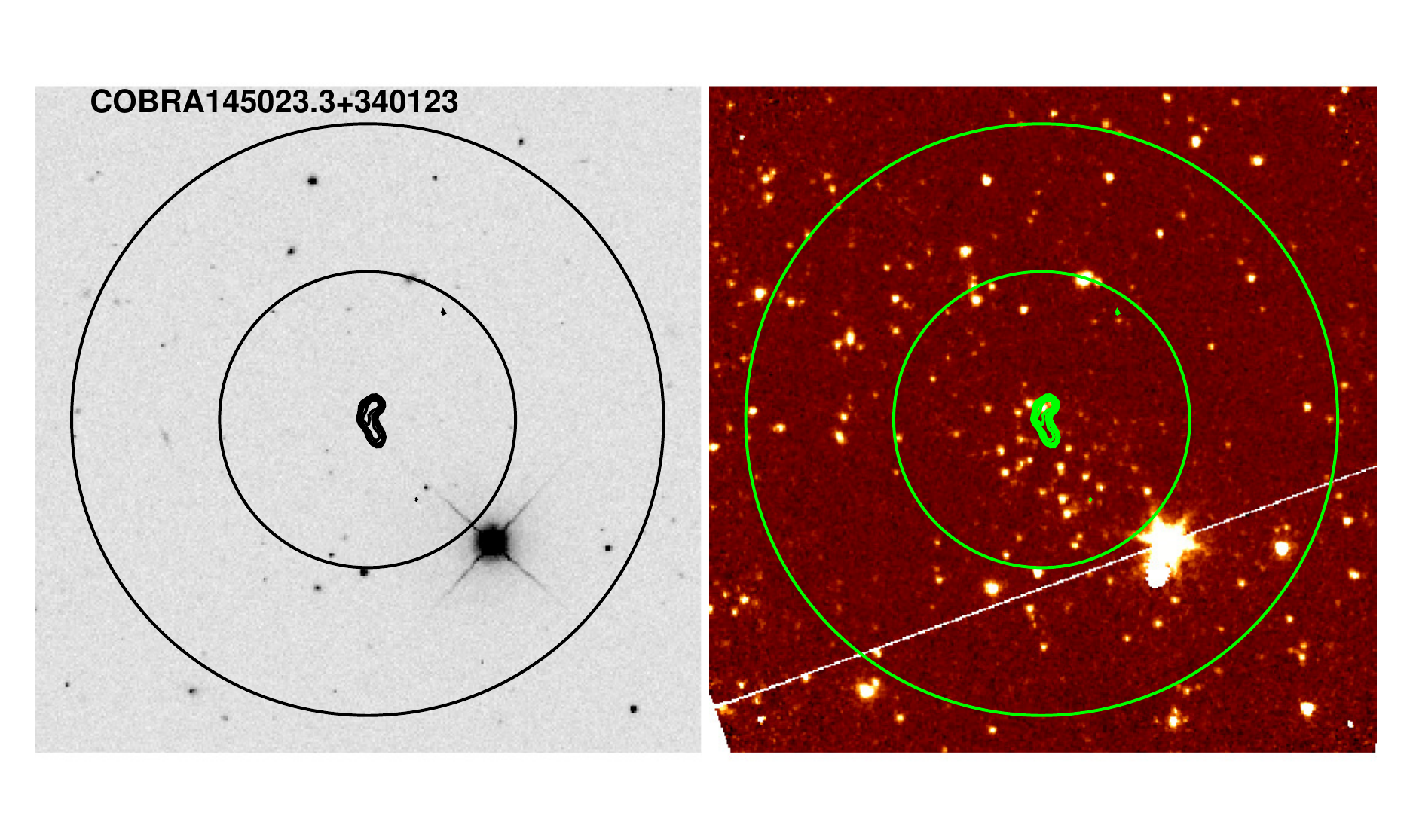}}
\subfigure{\includegraphics[scale=0.4, trim={0.0in 0.25in 0.0in 0.25in}, clip=true]{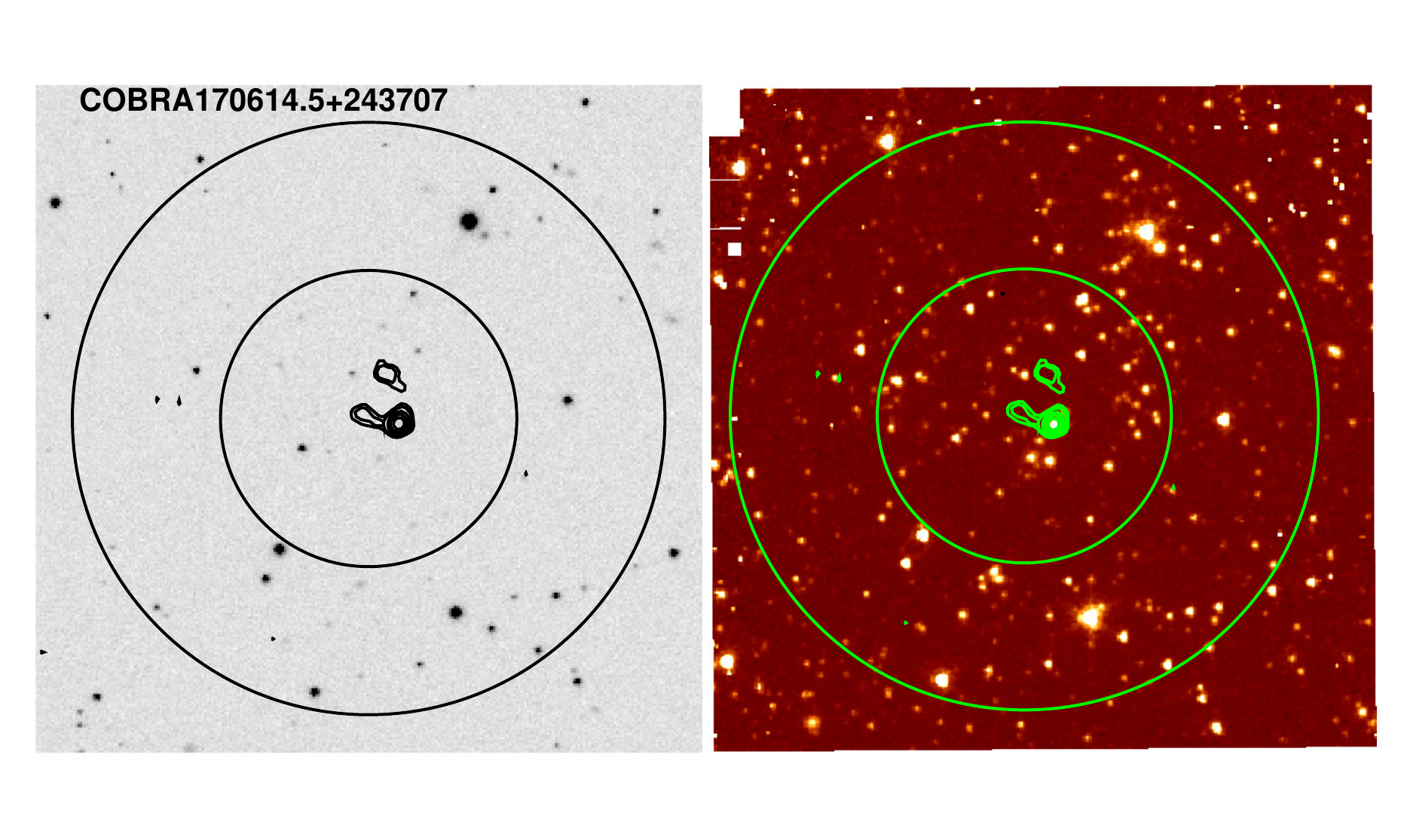}}
\subfigure{\includegraphics[scale=0.4, trim={0.0in 0.25in 0.0in 0.25in}, clip=true]{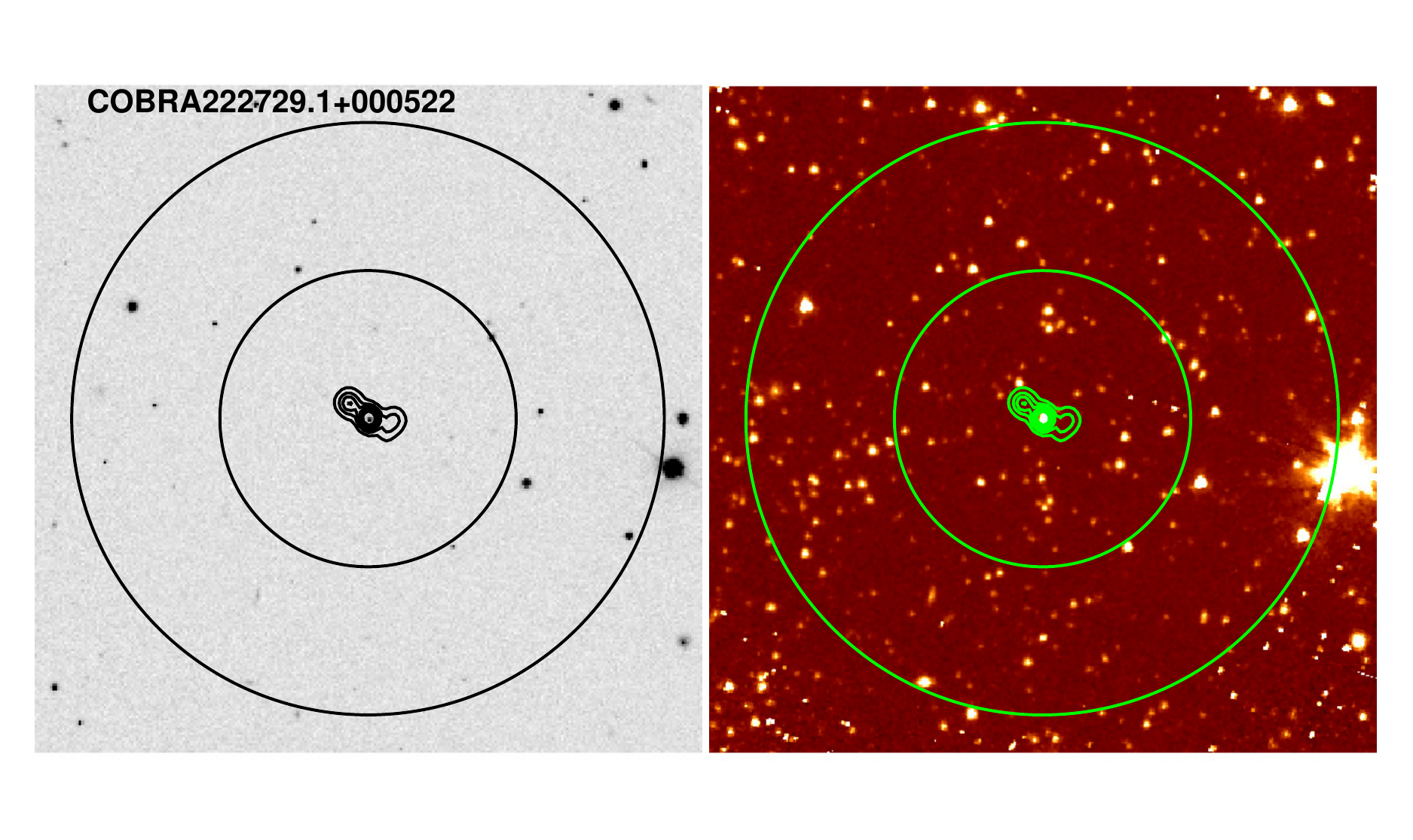}}
\caption{Montage showing the SDSS $r$-band and IRAC 3.6~\micron mosaics of six representative COBRA cluster candidates.  Greyscale: SDSS-r image of the area around the radio source.  Orange: \spitzer 3.6~\micron image.  The VLA FIRST radio contours are overlaid on both panels. The inner circle in both panels is one arcminute in radius, while the outer circle is two arcminutes in radius.  The radio host is not visible in the SDSS-r band, but becomes visible in the 3.6~\micron~\spitzer image, along with additional sources.  In all panels, north is up and east is to the left.}
\label{comparison}
\end{center}
\end{figure*}

We have selected six examples (Figure~\ref{comparison}) of our candidate clusters to discuss in more detail.  We will discuss one source for which a redshift could not be estimated, two with estimated but unconfirmed redshifts, one source observed in both bands that also has a spectroscopically confirmed redshift in SDSS, and two quasars, which have known redshifts.  All of the sources show clear bent morphology and have moderate-to-high overdensities.  For those sources with redshifts, we also discuss the extent of the radio source {(as measured from the edge of one lobe to the edge of the other)} and the radio luminosity of the source.  For the five of these six radio sources for which we have redshifts, we can calculate the radio luminosity of the source using the following equation:
\begin{equation}
L_{rad}=4\pi D_L^2 S_{\nu_0}\int_{\nu_1}^{\nu_2}\left(\frac{\nu}{\nu_0}\right)^{-\alpha}d\nu,
\label{radio}
\end{equation}
where $D_L$ is the luminosity distance, $S_{\nu_0}$ is the total flux density at the reference frequency $\nu_0$, and $\alpha$ is the spectral index.
For all our sources, we assume a spectral index $\alpha=0.8$, as is typical for extragalactic radio sources~\citep{sarazin}.  We take 1400~MHz as our reference frequency, corresponding to the average frequency where FIRST measured the radio flux densities for our sources, and use the integration limits of $10^7$~Hz to $10^{11}$~Hz, taking into account the redshift of each source.  The radio flux density for each source was measured at 20~cm by FIRST.

The candidate clusters in this survey, including those discussed below, are being { imaged in the optical} with the Discovery Channel Telescope (DCT) as part of a comprehensive follow-up program {to study these objects in more detail}.  The DCT is a 4.3~meter telescope located in Happy Jack, AZ.  These objects are being observed (Golden-Marx et al., in preparation) with the Large Monolithic Imager, a single chip CCD with a $12\farcm3\times 12\farcm3$ field of view.  

%
%

\textit{COBRA091928.9+273215:}  COBRA091928.9+273215 was observed in both bands, and is an example of a moderately overdense region.  It has an excess of 19.2 sources within 1\arcmin\ and an excess of 44.5 sources within 2\arcmin.  This corresponds to a Gaussian overdensity of $3.0\sigma$ within both radii.  This source has a photometric redshift of $z=0.63$ in the SDSS.  The radio lobes extend 50\farcs8, which corresponds to 347~kpc at this redshift.  Its radio luminosity is $L_{rad}=1.35\times10^{43}$~erg~s$^{-1}$, using a luminosity distance $D_L=3745.7$~Mpc and a 20~cm radio flux density of $S_0 = 52.9$~mJy. 


\textit{COBRA104254.8+290719:}  COBRA104254.8+290719 was observed in both bands.  It has an excess within 1\arcmin\ of 22.2 sources, which corresponds to a Gaussian significance of $3.5\sigma$.  It is not overdense within 2\arcmin.  Because we had observations in both bands, we used the color of the host to estimate a preliminary redshift, as discussed in \S\ref{zest}.  We find that this source has an approximate redshift of $z=1.04$.  It extends 54\farcs3, which corresponds to 438.7~kpc at this redshift.  Its radio luminosity is $L_{rad}=3.06\times10^{43}$~erg~s$^{-1}$, using a luminosity distance $D_L=6935.5$~Mpc and a 20~cm radio flux density of $S_0 = 33.3$~mJy.  This candidate has DCT follow-up observations, which show overdensities consistent with those observed with \spitzer (Golden-Marx et al., in preparation).


\textit{COBRA141155.2+341510:}  COBRA141155.2+341510 is a quasar with redshift of 1.82 and has an overdensity within both one and two arcminutes.  Within a 1\arcmin\ radius (506~kpc), it has an excess of 17.2 sources, which corresponds to 2.7$\sigma$ when compared to the Gaussian fit of the background.  Within a 2\arcmin\ radius the overdensity increases to 3.0$\sigma$, which corresponds to 44.5 sources above the background (the overdensity can be seen to the northwest of the radio source).  As mentioned in \S\ref{clustercan}, quasars may be triggered during infall, and thus reside on the outskirts of clusters.  It is possible that we are seeing this here.   While the quasar is visible in the SDSS-r image, the majority of the other sources are not.  The lobes of the quasar are 35\farcs8 (709~kpc) in extent.  The total radio luminosity is $L_{rad}=7.81\times10^{44}$~erg~s$^{-1}$, using a luminosity distance of $D_L=13840.7$~Mpc and a reference radio flux density of $S_0=200.1$~mJy.  This source has been observed with the DCT and shows an overdensity in the deep optical images, as well. 

\textit{COBRA145023.3+340123:}  COBRA145023.3+340123 is an example of an highly overdense target within 1\arcmin\ but not within 2\arcmin.  It was observed at 3.6~\micron only, although it has been observed at optical wavelengths with the DCT.  It has an excess of 23.2 sources within 1\arcmin\ which corresponds to an overdensity of $3.6\sigma$.  The extent of the radio source is 23\farcs2.  If we assume a redshift of $z=1.0$ (the approximate peak of the redshift distribution shown in Figure~\ref{zhist}), this corresponds to 185.8~kpc.   


\textit{COBRA170614.5+243707:}  COBRA170614.5+243707 is another highly overdense region in our sample.  It was observed in both bands, and has a redshift of $z=0.71$, which was confirmed spectroscopically in SDSS.  It has an excess of 29.2 sources within a 1\arcmin\ region, which corresponds to a significance of $4.6\sigma$.  Within 2\arcmin\ the excess is 72.5 sources, which corresponds to a Gaussian significance of $4.9\sigma$.  The radio lobes extend 46\arcsec (330.6~kpc).  The total radio luminosity of this source is $L_{rad}=6.37\times10^{42}$~erg~s$^{-1}$.  This was calculated using a luminosity distance $D_L=4334.1$~Mpc and a reference flux density $S_0=18.4$~mJy.  This source has been observed with the DCT.  


\textit{COBRA222729.1+000522:}  COBRA222729.1+000522 is a quasar with a redshift of 1.513. It is overdense within both 1\arcmin\ (508~kpc) and 2\arcmin.  {Within 1\arcmin\ there is an excess of 13.2 sources, which corresponds to a significance of 2.1$\sigma$.} Within 2\arcmin, there is an excess of 45.5 sources, which corresponds to a significance of 3.1$\sigma$.  As with J141155.2+341510, the stronger overdensity within 2\arcmin\ than 1\arcmin\ suggests that the quasar resides on the cluster outskirts.  The lobes of the quasar extend 30\farcs9 (261.5~kpc), and it has a radio luminosity $L_{rad}=4.32\times10^{44}$~erg~s$^{-1}$.  The radio luminosity was calculated using a luminosity distance $D_L=11025.2$~Mpc and a reference flux density $S_0=178.6$~mJy.  {This source has been observed with the DCT.} 

\section{Discussion and  Conclusions} \label{conc}

The high-redshift COBRA survey consists of 646 bent, double-lobed radio sources observed with \spitzer in the 3.6~\micron band (135 of these 646 sources were also observed at 4.5~\micron).  Forty-one of these sources are quasars with {spectroscopic} redshifts above $z=0.7$; the range extends to $z\approx3$.  We compared the number of sources as determined by running SExtractor on the region around the radio source to the expected number of sources based on the surface density of the SpUDS field.  We found 190 fields with overdensities corresponding to at least a $2\sigma$ significance.  These 190 over-dense regions are likely clusters, {most of them at high redshifts}.  Of the 190 cluster candidates, {39 have $z<0.5$}, 32 have $0.5\le z < 0.7$ and 119 have $z\ge0.7$ or do not have a known redshift, indicating that they are likely to have $z\ge0.7$.  {We will explore the relationship between overdensity and redshift in a forthcoming paper.}

\citet{w13} performed a similar analysis using obscured and unobscured powerful radio-loud (but not necessarily bent) AGN in the redshift range $1.2<z<3.2$ {(the Clusters Around Radio-Loud AGN, or CARLA, sample)} and found that 92\% of sources reside in a denser-than-average region.  They also find that the majority (55\%) of their sources reside in regions that are overdense at at least the $2\sigma$ level.  {To directly compare their results with ours, we ran their images through our pipeline, using the same parameters and methodology as described in \S\ref{sextract} and \S\ref{cluster}.  Doing so, we find that 43.9\% of the CARLA fields are overdense at the 2$\sigma$ level or higher and 88.1\% of the CARLA fields have a positive excess of galaxies as compared to the background.}   At a $2\sigma$ or higher confidence level, we find that 29.4\% of our {fields are overdense, while 82.0\% of our fields have a positive excess of galaxies as compared to the background.  Two sources, COBRA073320.4$+$272103 and COBRA143331.9$+$190711, appear in both surveys.  Both objects are quasars. COBRA073320.4$+$272103 is found to be very overdense ($>3.5\sigma$) in both surveys, while COBRA143331.9$+$190711 has a positive excess of galaxies in both surveys but is not overdense at the $\ge2\sigma$ level.  Both surveys are finding overdense regions at a similar rate.  The mean radio power of the sources in the CARLA sample is higher than in our COBRA sample.  These powerful, radio-loud AGN tend to be found in somewhat richer environments on average~\citep{hatch}.  Bent sources are found in a wide range of environments, including clusters, groups, and even large-scale filaments~\citep{edwards}.}

Using a somewhat different method, in another study, \citet{galametz} find that 73\% of their radio sources reside in regions with overdensities of 15 or more sources within one arcminute.  All of these surveys show that regardless of morphology, radio AGN are good tracers of high-redshift clusters. 

At low redshift, bent, double-lobed radio AGN are found in both relaxed (A2029, \citealp{clarke, rpm}) and merging (A562, \citealp{douglass}) clusters.  \citet{simpson} suggest that at redshifts $1<z<2$ radio AGN may be triggered by galaxy-galaxy mergers that are most likely triggered by cluster mergers.  As \citet{brodwin13} discuss, the merging causes a burst of star formation and fuels an AGN that eventually quenches the star formation by heating up the cold gas and/or expelling it.  Such merging may also explain the rapid mass growth observed by \citet{mancone}, the high AGN incidence observed by \citet{martini} and \citet{galametz09}, the scatter in cluster red sequences and young ages of red sequence galaxies observed by \citet{snyder}, as well as many other measurements.  These galaxy-galaxy mergers can also happen in lower-mass clusters whose smaller velocity dispersions lead to increased merger efficiency.  Such clusters need not be disturbed by cluster-scale mergers.  Recently \citet{cooke} identified a high-redshift ($z=1.58$) relaxed cluster selected using a powerful radio AGN as a signpost, indicating that such sources are found in a variety of environments at both high- and low-redshift.  AGN may also be triggered by inflowing gas in relaxed, cool core clusters{\citep{mcnamara}}.  Our sample will be sensitive to both merging and relaxed groups and clusters, and thus we will be able to explore galaxy evolution in both environments.  Combined with the wide range of masses we are probing, we will explore how these properties affect galaxy evolution in ongoing work.    
 
 \acknowledgements
 First and foremost, we thank the referee for a thorough and valuable review.
 
  We thank the CARLA team for generously sharing their data with us.
 
 RPM would like to thank Michael Malmrose for useful discussion.
  
This work has been supported by the National Science Foundation, grant AST-1309032.

This work is based in part on observations made with the \spitzer Space Telescope, which is operated by the Jet Propulsion Laboratory, California Institute of Technology under a contract with NASA. Support for this work was provided by NASA through an award issued by JPL/Caltech (NASA award RSA No. 1440385).

Funding for SDSS-III has been provided by the Alfred P. Sloan Foundation, the Participating Institutions, the National Science Foundation, and the U.S. Department of Energy Office of Science. The SDSS-III web site is http://www.sdss3.org/.

SDSS-III is managed by the Astrophysical Research Consortium for the Participating Institutions of the SDSS-III Collaboration including the University of Arizona, the Brazilian Participation Group, Brookhaven National Laboratory, Carnegie Mellon University, University of Florida, the French Participation Group, the German Participation Group, Harvard University, the Instituto de Astrofisica de Canarias, the Michigan State/Notre Dame/JINA Participation Group, Johns Hopkins University, Lawrence Berkeley National Laboratory, Max Planck Institute for Astrophysics, Max Planck Institute for Extraterrestrial Physics, New Mexico State University, New York University, Ohio State University, Pennsylvania State University, University of Portsmouth, Princeton University, the Spanish Participation Group, University of Tokyo, University of Utah, Vanderbilt University, University of Virginia, University of Washington, and Yale University.

\facilities{\spitzer, Sloan}

\bibliographystyle{apj}
\bibliography{cobra}

\end{document}